## Engineering Chirality in Halide Perovskites

*Juan Delgado-Alvarez,[1] Javier Castillo-Seoane,[1] Jorge Budagosky,[1,2] Lidia Contreras-Bernal,[1,3] Maria Alcaire,[1] Xabier Garcia-Casas,[1] José Feria,[1] Vanda Godinho,[1] Ana Borras,[1] Angel Barranco,[1]* Juan R. Sanchez-Valencia[1]**

1. Nanotechnology on Surfaces and Plasma Lab. Materials Science Institute of Seville. Consejo Superior de Investigaciones Científicas (CSIC)- Univ. Sevilla. c/ Américo Vespucio 49, 41092 Sevilla, Spain
2. Departamento de Física Aplicada I, Universidad de Sevilla, c/ Virgen de África 7, 41011, Seville (Spain)
3. Departamento de Química Física, Facultad de Química. Universidad de Sevilla. C/Profesor García González 41012, Seville (Spain).

E-mail: angelbar@icmse.csic.es, jrsanchez@icmse.csic.es

### Abstract

The ability to control chirality in halide perovskites offers new opportunities for circularly polarized photonics, spin-selective electronics, and quantum information technologies. Chirality in halide perovskites is commonly achieved through chiral molecular building blocks or externally imposed photonic architectures.[1–3] Although both approaches can generate strong chiroptical responses, many spin-dependent functionalities require structural symmetry breaking embedded within the material itself.[4–6] Here we show that chirality can emerge directly during crystal growth. By combining glancing angle deposition with controlled substrate rotation, we generate highly textured $PbI_2$ nanostructures with growth-controlled crystallographic torsion. X-ray texture analysis reveals that substrate rotation progressively rotates the crystal orientation during growth while preserving the c-axis orientation, resulting in a twisted texture with giant and tuneable chiroptical responses, including ellipticities of 19° and absorption dissymmetry factors approaching 0.6. The chirality programmed during growth is transferred through vapour-phase conversion into multiple halide perovskite compositions, resulting in circularly polarized luminescence with $g_{lum}$ values up to 0.23. These findings establish growth-controlled crystallographic torsion as a previously unexplored origin of chirality in halide perovskites.

## Main

Halide perovskites have emerged as a leading semiconductor platform owing to their outstanding optoelectronic properties, including strong light absorption, efficient light emission, long carrier diffusion lengths, and defect tolerance.[7,8] The introduction of chirality is further expanding their functionality, enabling materials that emit circularly polarized light, exhibit spin-selective charge transport, ferroelectricity, and strong nonlinear optical responses.[4–6,9–13] This convergence of optoelectronics and chirality promises disruptive advances in photonics and quantum information technologies. Over the past decade, several strategies have been developed to introduce chirality into halide perovskites, yet they rely on fundamentally different physical mechanisms.[1–3,14]

A first approach relies on externally imposed chirality. Photonic architectures and mesoscale templates can endow achiral perovskites with strong chiroptical activity, generating circular dichroism (CD) and circularly polarized luminescence (CPL) through chiral organization at length scales larger than the crystal lattice.[14–21] Helical metasurfaces, gammadion structures, cholesteric liquid crystals, cellulose nanocrystals, supramolecular gels, and electrospun fibres represent notable examples, with reported photoluminescence dissymmetry factors ($g_{lum}$) approaching 1.9.[17–21] Despite their impressive performance, the resulting chiral responses are determined by the surrounding architecture rather than by the intrinsic crystal structure of the perovskite.

A second strategy relies on chirality transfer from chiral molecular building blocks to the perovskite framework. In most reported systems, chirality originates from chiral organic cations that induce local structural distortions within the inorganic framework, including octahedral tilting, asymmetric bonding environments, surface chirality, and local inversion asymmetry.[2,22,23] Such distortions can twist Pb-X (X = halide) octahedra and transfer chirality to the electronic structure, thereby enhancing CD signals and even influencing spin splitting.[22,23] Modulating these interactions enhances chirality transfer and spin-polarization sensitivity in layered perovskites.[24,25] Moreover, the choice of chiral ligand dictates dimensionality, from 0D clusters to 3D frameworks, with lower-dimensional systems generally exhibiting stronger chiroptical responses due to quantum confinement.[2] However, this approach typically yields modest absorption dissymmetry factors ($g_{abs}$, the standard figure-of-merit for benchmarking chiroptical absorption), with chirality being primarily dictated by molecular composition rather than by controllable structural organization.

A particularly rare class of chiral halide perovskites exhibits genuine crystallographic helicity through enantiomorphic space groups containing chiral screw axes, generating non-superimposable crystal lattices with long-range helical order.[2] To date, only a few halide perovskites have been reported in such structures, including systems crystallizing in $P4_32_12/P4_12_12$[26] and $P6_1/P6_5$[13], where chirality is programmed directly in the crystallographic symmetry of the lattice and can produce $g_{abs}$ values approaching $10^{-2}$. Despite their remarkable properties, these materials remain uncommon because they require both chiral molecular building blocks and crystallization into rare helical space groups.

Glancing angle deposition (GLAD) is a physical vapor deposition strategy in which the vapor flux reaches the substrate at a highly oblique incidence angle.[27,28] This geometry promotes self-shadowing during growth, enabling the formation of tilted, columnar, and more complex nanostructured architectures with controlled anisotropy.[29,30] In its conventional implementation, substrate rotation during deposition can generate helical nanoarchitectures whose chiroptical properties arise from geometry rather than from the crystal structure itself.[29,30] Our previous work showed that $PbI_2$, which can subsequently be converted into halide perovskites through vapour-phase reactions, forms highly textured crystalline nanowalls with a nearly vertical orientation when deposited at fixed zenithal angles.[31] Similar behaviour has been reported for Mg and Zn nanostructures, where high surface diffusion dominates over self-shadowing effects during growth.[32,33] Importantly, these $PbI_2$ nanowalls develop a strong preferential crystallographic orientation that is preserved after conversion into halide perovskites, where it governs anisotropic optical absorption and photoluminescence.[31] These observations suggest that GLAD may provide a unique platform for controlling crystallographic orientation during growth, beyond its established role as a tool for engineering morphology.

If crystallographic orientation can be programmed during growth, can chirality also emerge from growth-controlled crystallographic torsion? Rather than relying on chiral molecules, rare helical space groups, or external photonic architectures, chirality may arise from the progressive rotation of crystal orientation during growth. While such an approach has not been explored in halide perovskites, twisted crystallites in other material systems have been shown to produce chiroptical responses through self-organization or strain-driven structural rotation.[15,34,35]

Here we show that controlled azimuthal rotation during GLAD enables a previously unexplored form of growth-encoded chirality in halide perovskites. By programming the evolution of crystallographic orientation during growth, $PbI_2$ nanostructures develop tuneable crystallographic torsion that persists after their conversion into halide perovskites. The resulting materials exhibit giant and tuneable chiroptical responses, with $g_{abs}$ approaching 0.6 and $g_{lum}$ values up to 0.23. More broadly, our results demonstrate that chirality can emerge as a growth-programmed property, establishing crystal growth as a new route for engineering chiral functionality in semiconductor materials.

**Growth-encoded chirality in $PbI_2$ nanostructures as perovskite precursors**

To explore whether chirality can emerge from growth-controlled crystallographic torsion, we extended our previously reported GLAD approach for the growth of highly textured $PbI_2$ nanowalls.[31] In that work, deposition at a fixed zenithal angle of 85° produced nearly vertical nanowalls with strong structural and optical anisotropy.[31] Importantly, $PbI_2$ serves as a versatile precursor that can subsequently be converted into multiple halide perovskite compositions while largely preserving the underlying nanostructure and crystallographic texture. Here, the zenithal angle ($\theta_D$) was kept constant at 85º, while the azimuthal angle ($\varphi_D$) was continuously rotated during deposition to induce a progressive evolution of crystallographic orientation throughout growth (further details required to replicate the rotation system are provided in **Supplementary Information S1**). **Figure 1a** illustrates the deposition configuration (see also Methods), in which substrate rotation is controlled by an automated stepper motor while maintaining a fixed deposition angle. Unlike conventional GLAD architectures, where chirality arises primarily from helical morphology,[29,30] the present approach seeks to generate chirality through growth-controlled crystallographic torsion. Consistent with this objective, scanning electron microscopy reveals twisted ribbon-like nanostructures rather than the helical columns typically associated with chiral GLAD geometries (**Fig. 1b**).[27,28]

To control the degree of crystallographic torsion, the substrate rotation speed was varied while maintaining a constant film thickness of 2.5 μm. Unless otherwise stated, all samples were grown using counterclockwise substrate rotation, which defines the left-handed ($L$-$PbI_2$) architecture. The resulting structures are described by their vertical pitch (VP), defined as the film thickness that would be deposited during one complete 360° substrate rotation. Samples spanning VP values from 1.7 to 26.7 μm were prepared, together with non-rotating (VP=∞) and rapidly rotating (VP=8 nm, hereafter denoted VP~0) control samples. As illustrated schematically in **Fig. 1c**, decreasing VP progressively increases the torsion of the ribbon-like nanoarchitecture, providing a simple and continuous parameter for programming the chiral structure during growth.

Circularly polarized absorbance measurements revealed a strong and non-monotonic dependence of the chiroptical response on VP. Since the twisted ribbons are built from highly anisotropic $PbI_2$ nanowalls, the measured circular dichroism may contain contributions from linear dichroism and linear birefringence (LDLB).[25,36–39] To isolate the genuine chiroptical component, spectra were acquired in both front-side (β = 0°) and back-side (β = 180°) configurations and analysed using the front-back averaging procedure described in Methods and **Supplementary Figure S2**. **Figure 1d** shows the absorbance spectra under left- and right-circularly polarized (LCP and RCP) illumination for all investigated pitches, measured in both

front- and back-side geometries. Whereas the non-rotating (VP=∞) and rapidly rotating (VP~0) control references exhibit negligible circular-polarization-dependent absorption, samples grown with intermediate VP values display progressively larger differences between left- and right-circularly polarized light, reaching a maximum for VP = 6.7 μm. Notably, despite their similarly weak chiroptical response, the VP=∞ and VP~0 controls exhibit markedly different absorbance profiles, with the rapidly rotating sample displaying broader spectral features around 550 and 650 nm (**Supplementary Fig. S3**). This observation suggests that high-speed substrate rotation modifies the structural organization of the $PbI_2$ nanostructures even when it does not generate a strong chiroptical response in the visible range.

The influence of LDLB contributions becomes immediately apparent when comparing the front- and back-side measurements. This effect is particularly pronounced for samples with VP=3.3-13.3 μm. In the front-side geometry (β = 0°), left-circularly polarized light (LCP) is absorbed more strongly across most of the visible spectrum, whereas the opposite trend is observed in the back-side configuration (β = 180°). However, the wavelength regions displaying the largest polarization contrast differ significantly between the two orientations. Thus, these observations indicate that a substantial fraction of the measured signal originates from anisotropic linear optical effects, highlighting the need to separate the genuine chiroptical response from LDLB contributions.

The extracted chiroptical component, represented by $\theta_{chirop}$ and $g_{abs\text{-}chirop}$ (**Fig. 1e,g**), reveals a highly tuneable response governed by VP. No measurable chiroptical signal is detected for the static control (VP=∞, see top-right inset in Fig. 1e). Upon substrate rotation (for example, VP = 26.7 μm), two distinct spectral features emerge: a negative band at near-infrared wavelengths (~800 nm) and a positive band at shorter wavelengths (~500 nm). As VP decreases from 26.7 to 6.7 μm, the short-wavelength positive feature remains largely fixed in spectral position, varying mainly in intensity. In contrast, the long-wavelength negative band progressively increases in magnitude and shifts towards shorter wavelengths. This evolution culminates at VP = 6.7 μm, where the chiroptical response reaches a maximum, with $\theta_{chirop}$ around 19° (Fig. 1e) and $g_{abs\text{-}chirop}$ approaching 0.6 (Fig. 1g). Notably, as it will be detailed later, this chiroptical response is largely preserved after conversion into halide perovskites, demonstrating that growth-controlled crystallographic torsion can be transferred from $PbI_2$ templates to functional semiconductor materials.

The magnitude of the effect is sufficiently large to be perceived directly by the naked eye. As shown in the inset of Fig. 1g, the VP = 6.7 μm sample exhibits a clear optical contrast when viewed through left- and right-circularly polarized filters. The darker appearance under LCP illumination is consistent with the stronger LCP absorption measured in the front-side configuration, providing a direct visual manifestation of the enhanced chiroptical response.

Further reduction of VP produces a qualitative change in the spectral response. At VP = 3.3 μm the overall signal decreases substantially, whereas at VP = 1.7 μm the spectrum reverses: the negative feature now appears at shorter wavelengths (~520 nm), while the positive band shifts to the near-infrared (~740 nm). As a result, the energy of the main chiroptical feature (for this handedness, $L$-$PbI_2$, the negative branch) can be tuned across a large fraction of the visible spectrum simply by varying VP (Fig. 1e, bottom inset). In the limit of very rapid substrate rotation (VP~0, see top-right inset in Fig. 1e), the chiroptical response collapses to $\theta_{chirop}$~0.3° and $g_{abs\text{-}chirop}$~$4\times10^{-3}$, revealing the existence of an optimum torsion regime for chiral amplification.

To assess the possible influence of anisotropic linear optical effects, the corresponding LDLB contribution ($\theta_{LDLB}$) is shown in **Supplementary Fig. S4**. The LDLB signal reaches its largest amplitude for VP = 3.3-13.3 μm and becomes negligible for the static control (VP=∞), partially mirroring the trend observed for the chiroptical response. However, the spectral profiles differ markedly. In particular, the LDLB spectra do not exhibit the long-wavelength sign inversion observed in $\theta_{chirop}$, indicating that the two contributions arise from distinct physical mechanisms.[36,37] Although front-back averaging removes the dominant second-order LDLB contribution from the measured signal, higher-order anisotropic terms may persist after this correction. Previous studies have shown that such contributions are generally weaker than LDLB,[36,37] suggesting that, while residual higher-order anisotropic effects cannot be completely excluded, they are unlikely to alter the main conclusions derived from the chiroptical spectra. Furthermore, a detailed analysis presented in **Supplementary Information S5** demonstrates that the observed chiroptical response cannot be explained by circular Bragg reflection, despite the presence of a well-defined structural pitch.[29] Together, these observations support the conclusion that the giant optical activity originates from growth-induced chiral organization rather than from linear anisotropy or morphology-dependent photonic effects.

To further verify the chiral origin of the response, $PbI_2$ nanostructures were grown using counterclockwise and clockwise substrate rotations, producing left- ($L$-$PbI_2$) and right-handed ($R$-$PbI_2$) architectures, respectively. As shown in **Fig. 1f**, reversing the rotation direction leads to a nearly mirror-symmetric inversion of both the $\theta_{chirop}$ and $g_{abs\text{-}chirop}$ spectra, while preserving their spectral shape and magnitude. The corresponding front- and back-side absorbance spectra are presented in **Supplementary Fig. S6**, whereas the LDLB contribution is shown separately in Supplementary Fig. S4. This behaviour closely parallels the response observed for opposite molecular enantiomers in conventional chiral perovskites and provides direct evidence that the handedness of the chiroptical signal is determined by the growth process. The ability to reproducibly invert the optical response through substrate rotation alone demonstrates that chirality is encoded during growth rather than arising from measurement artefacts. More broadly, our GLAD approach enables direct control over both the handedness and the degree of crystallographic torsion through simple growth parameters such as the rotation direction and pitch. Combined with the subsequent vapour-phase conversion of $PbI_2$ into different perovskite compositions, this strategy provides an unusual degree of flexibility for independently engineering chirality and semiconductor functionality.

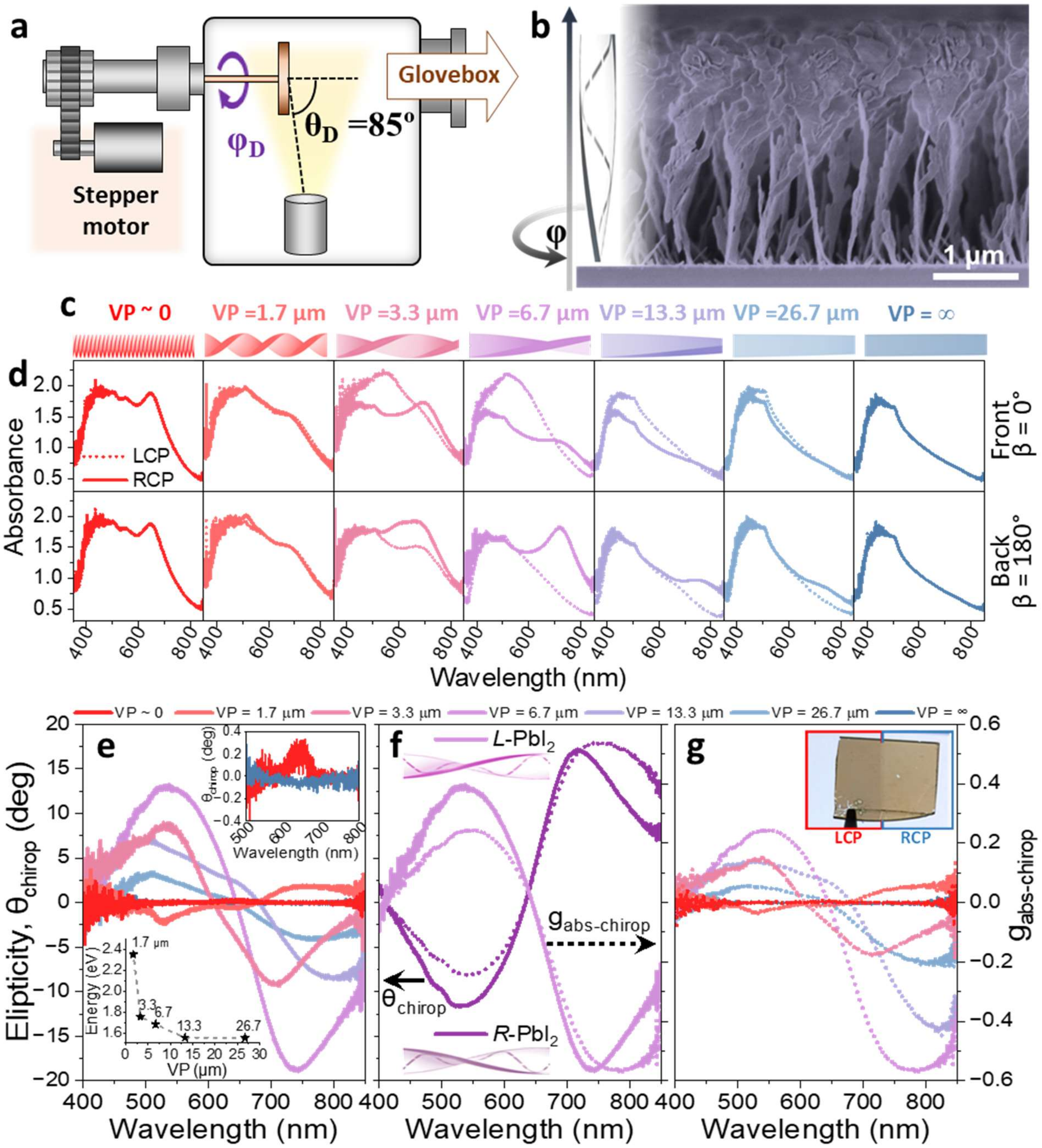

**Figure 1 | Growth-encoded chirality in GLAD $PbI_2$ nanostructures. a**, Schematic of the GLAD setup with a fixed zenithal angle $\theta_D$=85º and motor-controlled azimuthal rotation, $\varphi_D$. **b**, Cross-sectional SEM image of a *L*-$PbI_2$ film grown with VP=6.7 µm and a thickness of 2.5 µm. **c**, Schematic illustrating the pitch-dependent twisting of the ribbons. **d**, Absorbance spectra under left and right circularly polarized light (LCP and RCP) for L-$PbI_2$ samples with VPs from 1.7 to 26.7 µm, measured in both β=0º (front-side) and β=180º (back-side) orientations. Controls with no rotation (VP=∞) and high-speed rotation (VP~0) are included. **e, g**, Chiroptical ellipticity, $\theta_{chirop}$ (**e**), and dissymmetry factor, $g_{abs\text{-}chirop}$ (**g**), of *L*-$PbI_2$ samples presented in **c**, calculated using the front-back averaging method. Insets in **e** show (top right) a magnified view of $\theta_{chirop}$ of the VP=∞ and VP ~0 controls and (bottom left) the energy of the $\theta_{chirop}$ minimum (negative branch) as a function of VP. The inset in **g** shows a photograph of the VP = 6.7 µm sample viewed through LCP and RCP polarizers. **f**, $\theta_{chirop}$ and $g_{abs\text{-}chirop}$ for $PbI_2$ films grown with VP = 6.7 µm using counterclockwise (*L*-$PbI_2$) and clockwise (*R*-$PbI_2$) substrate rotation, producing left- and right-handed nanostructures, respectively.

**Growth-controlled crystallographic torsion.**

We begin by examining the static reference sample (VP=∞), whose (0003) pole figure and reconstructed ODF are shown in **Fig. 2b,c**. These measurements establish the baseline crystallographic texture in the absence of substrate rotation. In hexagonal $PbI_2$, the (0003) reflection directly probes the crystallographic c-axis orientation. Thus, in the pole figure representation, $\varphi_{PF}$ and $\psi_{PF}$ describe the azimuthal and tilt distribution of the c-axis, respectively. The texture is dominated by two sharp maxima located at $\varphi_{PF} \approx 90°$ and 270° and $\psi_{PF} \approx 75°$, consistent with the nearly vertical nanowall morphology previously reported for static GLAD growth.[31] Their finite angular extent reflects the intrinsic distribution of nanowall tilt angles, whereas the asymmetry in intensity arises from the directional nature of the oblique vapour flux. More importantly, the absence of additional maxima indicates that crystal growth selects a unique azimuthal orientation, a conclusion further supported by previous TEM observations of preferential (000l) diffraction from detached nanowalls.[31]

The ODF reconstructed from the three principal $PbI_2$ pole figures, corresponding to the (0003), (10-11) and (11-20) reflections, further reinforces this picture. Rather than exhibiting a broad distribution of crystallographic orientations, the texture is concentrated around a single c-axis tilt and a well-defined in-plane direction, manifested by two intensity cylinders along $\varphi_2$ (the rotation about the crystal c-axis), at $\varphi_1 \approx 0°$ and 180° and centred at $\Phi \approx 75°$ (the angle between the crystal c-axis and the film normal). The narrow extent of the ODF along $\varphi_1$ (the in-plane azimuthal orientation) confirms that the crystallographic orientation remains confined to a unique growth-selected azimuth. Together, the pole figures and ODF provide a self-consistent description of a strongly textured $PbI_2$ architecture whose crystallographic orientation is controlled by the directional deposition flux.

Because this orientational selection is imposed by the deposition geometry, substrate rotation during GLAD is expected to transfer the evolving flux azimuth to the growing nanowalls, producing a progressive rotation of their crystallographic orientation. This mechanism, schematically illustrated in **Fig. 2a**, would generate a twisted crystallographic texture while preserving the c-axis tilt, constituting a form of growth-controlled crystallographic torsion.

To test this hypothesis, we analysed pole figures from samples grown under substrate rotation corresponding to VP = 13.3 (**Figs. 2d-e**) and 6.7 μm (**Fig. 2f-g**). Relative to the static reference, the (0003) pole figure retains the maxima at $\psi_{PF} \approx 75°$, indicating that the $PbI_2$ c-axis remains strongly tilted during growth. However, the sharp maxima observed for VP=∞ progressively broaden as VP decreases. This evolution is already evident for VP = 13.3 μm and culminates at VP = 6.7 μm, where the intensity forms an almost continuous ring-like distribution at $\psi_{PF} \approx 75°$, instead of being concentrated around two discrete azimuthal directions.

The intensity of the ODFs remains centred at $\Phi \approx 75°$, indicating that the magnitude of the c-axis tilt is largely unaffected by substrate rotation. In contrast, the azimuthal distribution evolves dramatically with decreasing VP. The discrete intensity cylinders observed for the static sample broaden for VP = 13.3 µm and develop into a nearly continuous wall-like distribution for VP = 6.7 µm. This behaviour demonstrates that substrate rotation does not alter how much the c-axis is tilted, since the maxima appear at $\Phi \approx 75°$, but progressively rotates its in-plane orientation during growth. The broadening of the $\varphi_1$ distribution therefore provides direct evidence of growth-controlled crystallographic torsion, whereby the rotating deposition flux is transferred to the crystallographic orientation of the growing $PbI_2$ nanowalls.

The crystallographic torsion revealed by this XRD pole figure analysis provides an instructive comparison with previously reported helical halide perovskites. In those systems, helicity is imposed by crystallographic symmetry and therefore constrained to nanometric pitches. For example, 0D tin bromide clusters crystallizing in $P6_1/P6_5$ exhibit pitches of ~7.7 nm per full rotation and $g_{abs}$ values approaching $3.5 \times 10^{-2}$,[13] whereas 2D layered perovskites crystallizing in $P4_32_12/P4_12_12$ display pitches of ~4 nm and $g_{abs}$ values of ~$3 \times 10^{-4}$.[40] In our most comparable configuration, corresponding to the rapidly rotating limit (labelled as VP~0), the $PbI_2$ nanostructures exhibit a crystallographic twist of approximately 8 nm per full rotation and generate a modest but measurable chiroptical response ($\theta_{chirop}$~250 mdeg, and $g_{abs\text{-}chirop}$~$4\times 10^{-3}$, see **Supplementary Fig. S7).** These values are consistent with the range reported for symmetry-driven helicoidal lattices, despite arising from an entirely different mechanism.

Unlike symmetry-locked helices, however, growth-controlled crystallographic torsion is not restricted to nanometric pitches. By continuously decoupling the crystallographic rotation from the constraints imposed by the space group, our GLAD approach enables torsional pitches to be extended from nanometres to micrometres, where the strongest chiroptical responses are observed (Fig. 1). This additional degree of freedom provides a new design principle for chiral semiconductor materials and, as shown in the following section, remains largely preserved after conversion of $PbI_2$ into different halide perovskite compositions.

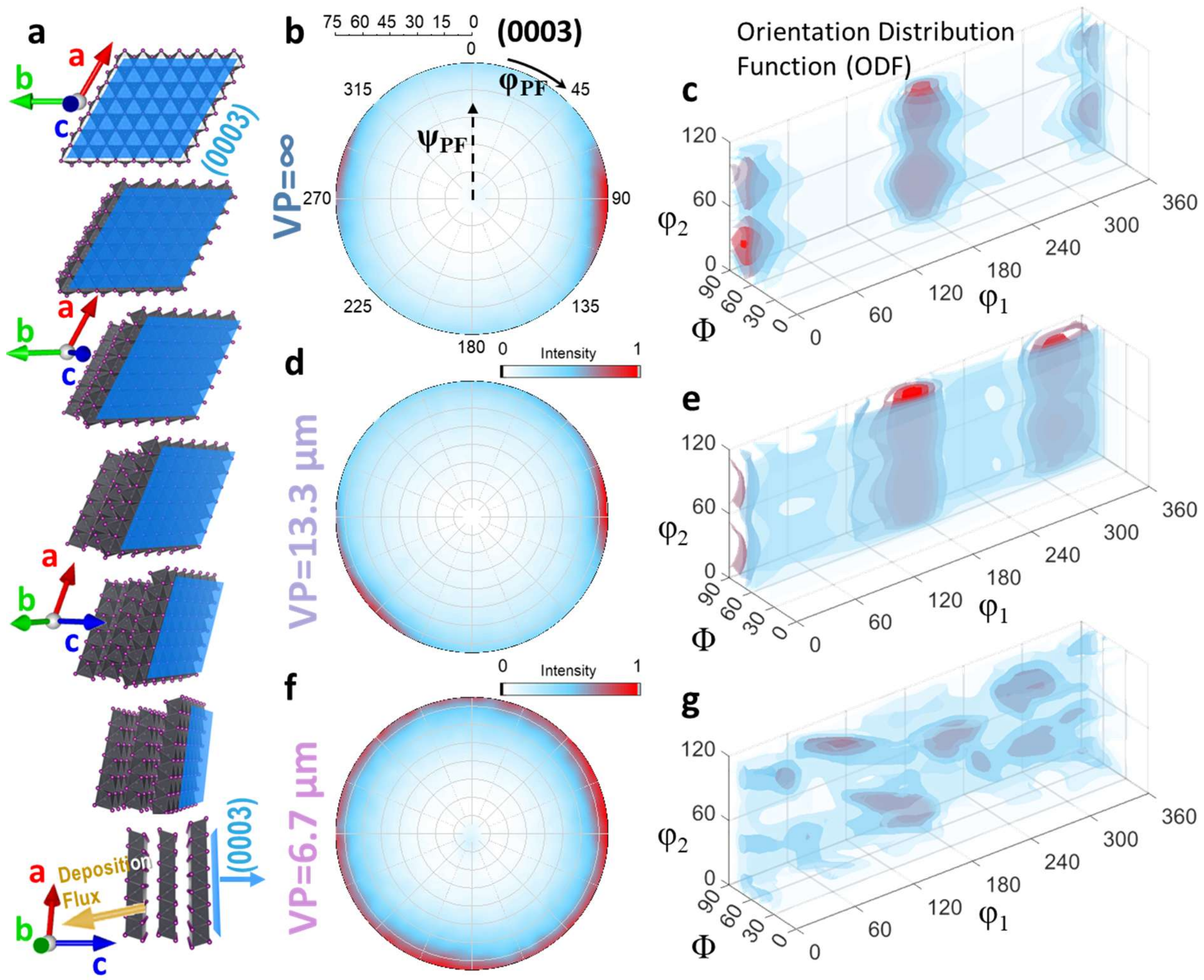

**Figure 2 | Growth-controlled crystallographic torsion in GLAD $PbI_2$ nanostructures. a.** Schematic of the proposed growth mechanism: during glancing-angle deposition, substrate rotation preserves the c-axis tilt (0003) while progressively rotating its in-plane azimuth, generating a twisted crystallographic texture. **b-g.** Pole figures of the (0003) (**b, d, f**) reflection of $PbI_2$, together with the corresponding orientation distribution functions (ODFs) reconstructed from the pole figures (**c, e, g**). Data are shown for a static film (VP =∞, **b-c**), and for films grown under substrate rotation corresponding to VP = 13.3 μm (**d-e**) and VP = 6.7 μm (**f-g**).

## Transfer of growth-encoded chirality to halide perovskites.

Having established that substrate rotation produces growth-controlled crystallographic torsion in $PbI_2$, we will show how the imprinted chirality is transferred to the halide perovskite by a vapor phase transformation. The small dimensions and high porosity of the $PbI_2$ nanostructures enable efficient vapour-phase transformation upon exposure to a secondary precursor while preserving the underlying nanoarchitecture.[31] The perovskite transformation is also achievable by closed-space sublimation inside a glovebox (used for bromine-based precursors to prevent cross-contamination, see Methods). Unlike liquid-phase conversion routes, which induce partial collapse and flattening of the nanowalls (**Supplementary Fig. S8**), vapour-phase processing maintains the structural features required to sustain the chiroptical response.[31]

**Figure 3a-b** shows the absorbance spectra (front and back side, left- and right-handed circularly polarized light illumination) of $PbI_2$ (yellow) and $MAPbI_3$ (blue) of left- and right-handed $MAPbI_3$ nanostructures obtained from MAI exposure of the corresponding *L*-$PbI_2$ and *R*-$PbI_2$ samples, respectively. Successful conversion to the corresponding halide perovskite is evidenced by the characteristic colour change from yellow to dark brown (Fig. 3a inset) and by the XRD analysis (**Supplementary Fig. S9**). Despite the complete chemical conversion from $PbI_2$ to $MAPbI_3$, both $\theta_{chirop}$ and $g_{abs\text{-}chirop}$ shown in Fig. 3b retain the characteristic spectral features of the parent nanostructures, exhibiting only minor red shifts after conversion. More importantly, although the absorbance spectra differ between the *L*- and *R*-samples, both $\theta_{chirop}$ and $g_{abs\text{-}chirop}$ display nearly mirror-symmetric profiles with opposite signs and comparable magnitudes, as expected for enantiomeric architectures. This behaviour demonstrates that the handedness programmed during $PbI_2$ growth is preserved throughout the perovskite transformation.

Beyond circularly polarized absorption, the converted perovskites exhibit strong circularly polarized luminescence (CPL). **Figure 3c** compares the CPL spectra of *L*- and *R*-$MAPbI_3$ under 625 nm monochromatic natural light excitation. The *L*-$MAPbI_3$ exhibits preferential right-handed CPL emission ($g_{lum} \approx -0.22$), while the *R*-$MAPbI_3$ analogue shows opposite handedness with similar magnitude ($g_{lum} \approx +0.23$), as expected for the two different rotations. These results demonstrate that the structural chirality programmed during growth determines not only the circularly polarized absorption but also the luminescence properties of the resulting perovskite.

The transfer of growth-encoded chirality is not limited to $MAPbI_3$. The same strategy can be extended to a broad family of halide perovskites, enabling independent tuning of the bandgap, composition, and optoelectronic properties while preserving both the nanostructure morphology and the chiroptical response. Representative examples are shown in **Fig. 3d-k** for $FAPbI_3$, $CsPbI_3$, $MAPbI_2Br$, and $FAPbI_2Br$ obtained by vapor exposure to FAI, CsI, MABr, and FABr, respectively, demonstrating that strong chiroptical activity is preserved across all compositions after conversion. A systematic CPL response is likewise observed across the entire perovskite materials family. *L*-handed perovskites consistently exhibit preferential right-handed CPL emission, reproducing the behaviour observed for $MAPbI_3$ (Fig. 3a-c). The corresponding $g_{lum}$ values for the *L*-enantiomeric architectures range from -0.13 to -0.23, with the highest values measured for $MAPbI_3$ (-0.23) and $FAPbI_3$ (-0.21). Moreover, $MAPbI_3$, $FAPbI_3$, and $CsPbI_3$ were synthesized in both *L*- and *R*-enantiomeric forms, yielding nearly identical absolute $g_{lum}$ values but opposite signs, further confirming that the CPL response originates from the handedness programmed during growth rather than from composition-dependent effects.

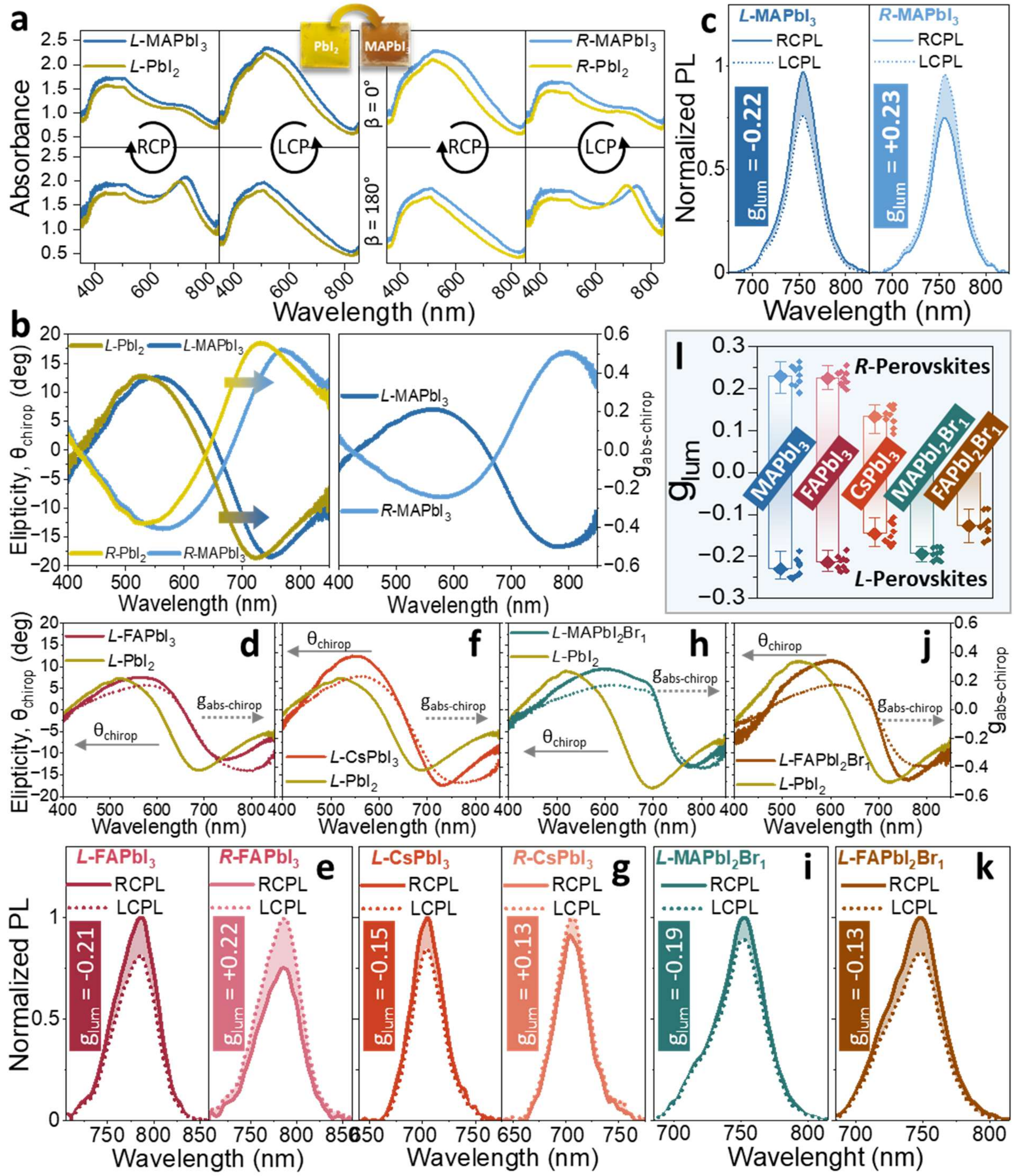

**Figure 3 | Transfer of growth-encoded chirality to halide perovskites.** **a**, Absorbance spectra under left- and right-circularly polarized light (LCP and RCP) measured in front- (β=0º) and back-side (β= 180º) geometries for $PbI_2$ nanostructures before conversion (yellow) and $MAPbI_3$ (blue) nanostructures after vapour-phase transformation. Data are shown for left-handed (L-$PbI_2$, counterclockwise rotation) and right-handed (R-$PbI_2$, clockwise rotation) architectures. Inset, photograph of the sample before and after exposure to MAI vapour. **b**, Corresponding chiroptical ellipticity, $\theta_{chirop}$ (left), and dissymmetry factor, $g_{abs\text{-}chirop}$ (right), calculated using the front-back averaging method from curves in panel **a**. **c**, Circularly polarized luminescence (CPL) of *L*- and *R*-$MAPbI_3$ samples under 625 nm monochromatic natural-light excitation. **d-k**, $\theta_{chirop}$ and $g_{abs\text{-}chirop}$ (**d, f, h, j**), together with CPL spectra (**e, g, i, k**), for perovskites derived from chiral $PbI_2$ templates and converted to $FAPbI_3$ (**d, e**), $CsPbI_3$ (**f, g**), $FAPbI_2Br$ (**h, i**), and $MAPbI_2Br$ (**j, k**). *L*- and *R*-enantiomeric architectures are shown where available. **l**, Summary of the $g_{lum}$ values obtained for all chiral perovskite compositions investigated.

**Figure 3l** summarizes the CPL performance of all investigated perovskites. Maximum $g_{lum}$ values approaching 0.23 are obtained, while all converted perovskites retain exceptionally large absorption dissymmetry factors, with $g_{abs}$ exceeding 0.4 and reaching values far above those typically associated with intrinsically chiral perovskites.[2,9–11] Collectively, these results demonstrate that giant chiroptical activity can emerge from growth-controlled crystallographic torsion without the need for chiral molecular ligands, chirality transfer mechanisms, or helicoidal crystal lattices. Once programmed during $PbI_2$ growth, this symmetry breaking can be transferred across multiple perovskite compositions while preserving both absorption and emission asymmetry. Beyond chiroptical activity, the ability to engineer chirality through crystal growth offers a previously unexplored route to control symmetry-breaking phenomena in semiconductor materials, independent of their chemical composition. This approach opens new opportunities for tailoring spin-dependent transport, ferroelectric, nonlinear optical and other chirality-enabled functionalities, establishing crystal growth as a powerful platform for the design of next-generation quantum, photonic and information technologies.

## Methods

***Precursors***. Lead iodide ($PbI_2$, 99.99% purity) was purchased from Merck. Methylammonium iodide (MAI), methylammonium bromide (MABr), formamidinium iodide (FAI), formamidinium bromide (FABr), and cesium iodide (CsI) were obtained from TCI with low water content and a purity of ≥98.0%, and were used as received without further purification.

***$PbI_2$ by Glancing Angle Deposition***. $PbI_2$ was sublimated using a low-temperature evaporator heated to approximately 330-350 °C. A schematic of the experimental setup is shown in Figure 1a. The distance between the evaporator and the substrate was 20 cm. $PbI_2$ deposition was performed at pressures below $5 \times 10^{-6}$ mbar onto fused silica and Si(100) wafer pieces. Substrates were mounted in a vertical position, and the zenithal deposition angle ($\theta_D$) was defined by the lateral displacement from the centre (≈1.8 cm for an angle of 85°). Substrates were mounted on a sample holder coupled to a magnetic transfer bar that maintained the specified displacement, provided angular control, and enabled transfer into a glovebox. Azimuthal rotation ($\varphi_D$) was controlled by a stepper motor controlled by ESP32-based electronics. Further details on the software, materials, and motor configuration necessary to replicate our system are provided in Supplementary Information S1.

The deposition rate was monitored using a quartz crystal microbalance (QCM) placed at the same height as the samples facing the evaporator, with its surface normal aligned with the incoming deposition flux, unlike the samples, which were tilted at a glancing angle. A constant growth rate of 0.6 Å/s was maintained throughout the process. While the QCM indicated a nominal thickness of 600 nm, cross-sectional SEM measurements revealed an actual thickness of approximately 2500 nm. During deposition, the samples were continuously rotated, and the total rotation defined the vertical pitch (VP). Samples with VP values of 1.7, 3.3, 6.7, 13.3, and 26.7 µm corresponded to swept angles (counterclockwise rotation, left-handed) of 540°, 270°, 135°, 62.5°, and 31.3°, respectively. For comparison, two additional samples were prepared: one with VP=∞ (static, no rotation) and another with VP ≈ 0 (maximum rotation speed, two full turns per minute), or, more precisely, VP=8 nm. Further samples were rotated clockwise (right-handed) to generate nanostructures with the opposite handedness.

***Transformation to halide perovskite.*** The conversion of $PbI_2$ nanostructures into halide perovskite was carried out by exposure to a second precursor under different conditions: for iodine salts, the transformation was done under high vacuum, as previously reported for $MAPbI_3$ nanowalls (VP=∞);[31] for bromine precursors, the conversion was performed by closed-space sublimation inside a glovebox (See **Supplementary Figure S10**).

For the vacuum transformation using iodine precursors (MAI, FAI, and CsI), the $PbI_2$ chiral samples were placed at 0º with respect to the evaporation source (non-glancing conditions, using a different sample holder). We observed that the deposition of MAI and FAI could not be properly controlled by the QCM. The MAI and FAI sublimation produces an increase in the pressure of the system. Based on these observations, we proceed to an optimization process and found reproducible and stable conditions by maintaining the evaporation temperature of the MAI and FAI precursors at 170ºC, which produces an increase in the pressure to around (5-6) $x10^{-5}$ mbar. Then, we reduced the pumping flux of the turbomolecular pump using a butterfly valve and fixed the pressure to $1 \times 10^{-4}$ mbar. By contrast, the CsI evaporation temperature was 380 °C with a rate of 0.1-0.2 Å/s and a final nominal thickness of 50 nm. The evaporation of CsI was carried out at a base pressure of $1 \times 10^{-5}$ mbar.

The conversion using bromine-based precursors (MABr and FABr) was carried out by closed-space sublimation in a custom-built setup (see Figure S10). This system consisted of two aluminum Petri dishes. The samples were placed on the upper dish, while the precursor powder was positioned on the lower one. The two Petri dishes were sealed with a PDMS gasket, and the upper dish was connected to a rotary pump through a tube, enabling operation under a rough vacuum of approximately 1 mbar and promoting uniform precursor transport during sublimation. The entire assembly was placed on a heating plate and heated to 150 °C for MABr and 160 °C for FABr. To prevent uncontrolled heating of the samples, the upper dish was cooled using a copper heat sink and a fan. Several photographs of the homemade closed-space sublimation system are shown in Figure S10.

It is worth noting that while $PbI_2$ can tolerate exposure to ambient conditions (moisture and oxygen), the synthesized perovskites cannot, due to the inherent sensitivity of hybrid halide perovskite materials, enhanced by the high porosity of the chiral nanostructures. Thus, all the optical and PL measurements have been performed inside the glovebox as detailed below.

***Characterization Methods.***

High-resolution scanning electron microscopy (SEM) images of the samples were acquired using a Hitachi S4800 microscope operating at 2 kV. Cross-sectional views were obtained by cleaving the Si substrates.

Pole-figure measurements were carried out using a Malvern Panalytical Empyrean X-ray diffractometer equipped with a five-axis Eulerian cradle, enabling independent variation of the sample tilt ($\psi_{PF}$) and azimuthal rotation ($\varphi_{PF}$). Cu Kα radiation ($\lambda$ = 1.5418 Å) was employed throughout the measurements. Pole figures were acquired from the (0003), (11-20), and (10-11) reflections of hexagonal $PbI_2$ by fixing the diffraction angle (2θ) at the corresponding Bragg positions. The tilt angle $\psi_{PF}$ was scanned from 0° to 75°, while $\varphi_{PF}$ was varied over the full 0°-360° azimuthal range using 5° angular increments. Raw pole-figure data were corrected for background and defocusing effects prior to texture analysis. Orientation distribution functions (ODFs) were reconstructed from the corrected pole figures using the MTEX toolbox,[41,42] adopting the Bunge Euler angle convention.

Circularly-polarized dependent UV-Vis-NIR spectroscopy measurements were carried out inside a glovebox ($O_2$ and $H_2O$ concentrations below 1 ppm) using a fibre-coupled custom optical setup. Illumination and transmitted light were guided by optical fibres connected to a tungsten halogen lamp light source (HL-2000-FHSA) and a UV-Vis CCD detector (Flame-S-UV-Vis-ES spectrometer from OceanOptics), both located outside the glovebox and interfaced through optical feedthroughs. The fibres were aligned using lenses acting as collimators. The collimated beam passed through a linear polarizer (Thorlabs GTH10M-A) and a rotatable quarter-wave plate (Thorlabs AQWP05M-600). Both components operate in the 400-800 nm range. The polarizer was mounted on a rotatable stage, enabling precise control of the angle between the polarization direction and the fast axis of the waveplate. This angle was defined as positive when measured clockwise from the polarization vector to the fast axis. Under this convention, right-handed circularly polarized light (RCP) is generated at +45°, while LCP is obtained at -45°. The sample was placed in the beam path using a rotatable holder, allowing measurements at different azimuthal orientations. Finally, the transmitted light was collected by a second collimating lens and guided through an optical fibre to the detector (see Figure S10).

Circularly polarized optical parameters, such as ellipticity (θ) and the absorption dissymmetry factor ($g_{abs}$), are key descriptors for quantifying chiroptical properties. The measured ellipticity ($\theta_{obs}$) is defined as:

$$\theta_{obs}(degrees) = \left(\frac{180}{\pi}\right)\frac{e^{\Delta A\frac{ln10}{2}}-1}{e^{\Delta A\frac{ln10}{2}}+1} \quad \text{(Equation 1)}$$

where $\Delta A = A_L - A_R$, and $A_L$ and $A_R$ are the absorbances measured using LCP and RCP light, respectively. All measurements were performed with the sample rotated azimuthally at α=0° and 90° around the surface normal. For each polarization, the absorbance was computed as the average of the two orientations, i.e., $A_L = \frac{1}{2}[(A_L)_{\alpha=0^\circ} + (A_L)_{\alpha=90^\circ}]$ and $A_R = \frac{1}{2}[(A_R)_{\alpha=0^\circ} + (A_R)_{\alpha=90^\circ}]$, which effectively suppresses residual polarization artefacts from the optical components.

In most reports, this expression is simplified to: $\theta_{obs}(degrees) = \Delta A\left(\frac{Ln10}{4}\right)\left(\frac{180}{\pi}\right)$, assuming that $\Delta A \ll 1$. However, in this work we apply the full form given in Equation 1, since the approximation introduces a noticeable deviation for the samples with the highest CD.

Despite the simplicity of the approach to calculate the ellipticity, in anisotropic systems, such as oriented thin films or nanostructured materials, the measured $\theta_{obs}$ signal may include contributions not only from genuine intrinsic chirality (often termed "true θ" or more precisely $\theta_{chirop}$) but also from artifacts arising from linear dichroism (LD) and linear birefringence (LB), collectively known as the LDLB effect.[25,36–39] To suppress these second-order contributions arising from LDLB, spectra were recorded with the sample oriented in both front and back configurations, i.e., flipped at β=180° relative to the incident beam.[25,36–39] This front-back averaging procedure cancels the LDLB terms, which reverse sign upon flipping (see Figure S2), while preserving the invariant chiroptical contribution.

These steps collectively ensure that the measured CD signal accurately reflects the intrinsic chiroptical response of the sample.[25,36–39] We use the term 'chiroptic' instead of 'true' (as in other studies[25]) to avoid implying that CD signals from LDLB are artificial. These signals arise from real differential interactions with circularly polarized light, even in the absence of intrinsic chirality. The term 'chiroptic' better distinguishes optical activity due to intrinsic or molecular chirality, while acknowledging that anisotropy-induced effects also yield meaningful circular dichroism. The challenge lies not in the measurement but in misinterpreting such signals as purely chiral. Accurate interpretation through careful analysis and correction of anisotropic artifacts is essential for advancing chiral photonics, spin-optoelectronics, and polarization-sensitive technologies. Taking this into consideration, the measured ellipticity ($\theta_{obs}$) is:

$$\theta_{obs} = \theta_{chirop} + \frac{1}{2}(LD' \cdot LB - LD \cdot LB')$$

The first term corresponds to the intrinsic chiral ellipticity signal, whereas the second term represents the contribution from the LDLB effect. This latter signal is measured along an arbitrary axis in the laboratory frame, with the prime notation indicating a 45° rotation of that axis. Since the LDLB signal reverses when the sample is flipped by β=180° relative to the light propagation direction, the $\theta_{chirop}$ component can be isolated by calculating the average of the spectra recorded from opposite orientations (front and back, β=0 and 180º). See Supplementary Figure S2 for further details.

$$\theta_{chi} \quad = \frac{1}{2}\left[(\theta_{obs})_{\beta=0^\circ} + (\theta_{obs})_{\beta=180^\circ}\right]$$

While higher-order contributions (e.g., pairwise interaction terms) may persist, their influence is typically minor, and front-back averaging remains a robust and widely accepted approximation.[25,36–39]

From the $\theta_{chirop}$ values, the absorption dissymmetry factor ($g_{abs\text{-}chirop}$) was calculated using:

$$g_{abs-chiro} \quad = \left(\frac{\pi}{180}\right)\frac{\theta_{chirop}(deg)}{\left({}^{Ln(10)}/_{4}\right) \cdot A} \approx \frac{\theta_{chiro}\ (deg)}{32.98 \cdot A}$$

where A is the sample absorbance. Since the homemade setup does not allow an easy removal of the polarizing elements, $A$ was estimated by averaging all spectra acquired for $\theta_{chiro}$ determination, as follows:

$$A = \frac{1}{4}\left[\left(\frac{A_L + A_R}{2}\right)_{\substack{\alpha=0^{\circ} \\ \beta=0^{\circ}}} + \left(\frac{A_L + A_R}{2}\right)_{\substack{\alpha=90^{\circ} \\ \beta=0^{\circ}}} + \left(\frac{A_L + A_R}{2}\right)_{\substack{\alpha=0^{\circ} \\ \beta=180^{\circ}}} + \left(\frac{A_L + A_R}{2}\right)_{\substack{\alpha=90^{\circ} \\ \beta=180^{\circ}}}\right]$$

To validate this approximation, an additional measurement of unpolarized absorbance was performed for the sample with the highest θ (VP = 6.7 µm). The comparison, shown **in Figure S11**, indicates that the approximation deviates at longer wavelengths. Using the unpolarized absorbance yields higher $g_{abs\text{-}chirop}$ values than the averaged approach, confirming that the latter systematically underestimates $g_{abs}$. Consequently, all reported values should be considered conservative estimates.

Circularly Polarized Photoluminescence (CPL) measurements were performed using the same optical setup employed for UV-Vis characterization, with the following modifications. Instead of a broadband white lamp, illumination was provided by a LED source at 625 nm (and 505 nm for $CsPbI_3$) purchased from Thorlabs (models M625F2 and M505F3). The excitation light was unpolarized. To suppress direct LED reflections reaching the CCD detector, a long-pass filter with cut-off wavelength of 695 nm (and 665 nm for $CsPbI_3$) was placed in the detection path. During photoluminescence measurements, the sample was illuminated at an incidence angle of approximately 50° relative to the substrate normal, while the emitted PL was collected along the surface normal. The emitted beam passed sequentially through a quarter-wave plate and a linear polarizer, whose orientation was adjusted to select left- or right-handed CPL (±45°), following the same angular conventions as in the absorbance measurements. Figure S10 shows a picture and a scheme of the setup used for measuring CPL.

The photoluminescence dissymmetry factor ($g_{lum}$), a standard figure of merit for circularly polarized luminescence, quantifies the difference between left- and right-handed circularly polarized emission and is defined as:

$$g_{lum} = 2\frac{I_L - I_R}{I_L + I_R}$$

where $I_L$ and $I_R$ are the PL intensities of left- and right-handed CPL, respectively. To minimize the influence of residual anisotropic effects and sample orientation, $g_{lum}$ was determined at azimuthal angles ranging from 0° to 360° in 45° increments. The value reported for each sample corresponds to the mean of the nine individual measurements, while the error bars represent the standard deviation of the resulting glum distribution. Individual values obtained at each azimuthal angle are summarized in Fig. 3l.

### Supplementary Information

Supplementary Information is available.

### Acknowledgements

We thank the projects PID2022-143120OB-I00, PID2025-175485OB-I00, and PCI2024-153451, and the research network NextPVNet (RED2024-154178-T) funded by MCIN/AEI/10.13039/501100011033 and by "ERDF (FEDER) A way of making Europe", Fondos Nextgeneration EU and "Plan de Recuperación, Transformación y Resiliencia". JB acknowledges the "VII Plan Propio de Investigación y Transferencia" of the University of Seville. The authors also want to thank CSIC for its financial support through the Interdisciplinary Thematic Platform (PTI) Transición Energética Sostenible (PTI-TRANSENER+). We also thank the "Consejería de Universidad, Investigación e Innovación" of "Junta de Andalucía" through the project DGP_PIDI_2024_02239. Project ANGSTROM was selected in the Joint Transnational Call 2023 of M-ERA.NET 3, which is an EU-funded network of about 49 funding organisations (Horizon 2020 grant agreement No 958174). The project leading to this article has received funding from the EU H2020 program under grant agreement 851929 (ERC Starting Grant 3DScavengers).

# Supplementary Information

## Engineering Chirality in Halide Perovskites

*Juan Delgado-Alvarez,*[1] *Javier Castillo-Seoane,*[1] *Jorge Budagosky,*[1,2] *Lidia Contreras-Bernal,*[1,3] *Maria Alcaire,*[1] *Xabier Garcia-Casas,*[1] *José Feria,*[1] *Vanda Godinho,*[1] *Ana Borras,*[1] *Angel Barranco,*[1]* *Juan R. Sanchez-Valencia*[1]*

1. Nanotechnology on Surfaces and Plasma Lab. Materials Science Institute of Seville. Consejo Superior de Investigaciones Científicas (CSIC)- Univ. Sevilla. c/ Américo Vespucio 49, 41092 Sevilla, Spain
2. Departamento de Física Aplicada I, Universidad de Sevilla, c/ Virgen de África 7, 41011, Seville (Spain)
3. Departamento de Química Física, Facultad de Química. Universidad de Sevilla. C/Profesor García González 41012, Seville (Spain).

E-mail: angelbar@icmse.csic.es, jrsanchez@icmse.csic.es

**Supplementary Information S1. Software, materials, and motor configuration for the rotation system.**

The homemade rotation system was implemented using a stepper motor controlled by an ESP32-based microcontroller platform, which provides Bluetooth and Wi-Fi connectivity for remote rotation control. The rotation was achieved through a transfer bar that not only connects the vacuum synthesis chamber to the glovebox but also enables rotation of the samples. The connection between the transfer rod and the motor was accomplished by a synchronous GT2 rubber belt. The following list includes all the standard components required to replicate the system.

Components

- **Stepper motor NEMO 23 200 (SY57STH76-2006A).**
- **Controller TB6600**
- **Switching power supply, 24V, 6.5A.**
- **ESP32 (ESP32-WROOM-32D).**
- **Power supply, 5V, 1A.**
- **GT2 timing wheel (20 teeth, hole diameter 6.35 mm, 6 mm width)**
- **Rubber synchronous belt, 6 mm, GT2.** The length depends on the distance between the motor and the transfer bar (see description below).

In addition, the system requires several additional parts developed in our laboratory and provided in Supplementary Files S1 in a zip file, which contains CAD designed parts, PCB, and software required for the ESP32. The description is as follows.

1. **PCB board for controlling and calibrating the motor.** Figure S1-1 shows the PCB circuit layout (a–b) and a schematic view of its physical appearance (c). The PCB design files are provided in Supplementary Files S1 and can be used to replicate the board; these files can be submitted to any PCB fabrication service. The board provides a simple interface to connect the ESP32 output signals to the stepper motor. The ESP32 module is mounted on the female pin headers located on the left side of the PCB (see Figure S1-1b–c).

The female headers used on the PCB are longer than the length of the ESP32 module, leaving the uppermost pins extending beyond the footprint of the ESP32 (visible in the top region of Fig. S1-1b,c, where these two upper pins are labelled as NC (No Connect). Additionally, due to a design issue in this version of the PCB, the CMD pin of the ESP32 (corresponding to the second pin from the bottom in the left header row in Figures S1-1b,c) must be cut before installation, as leaving it connected causes a communication conflict with the board. This issue will be corrected in subsequent PCB revisions.

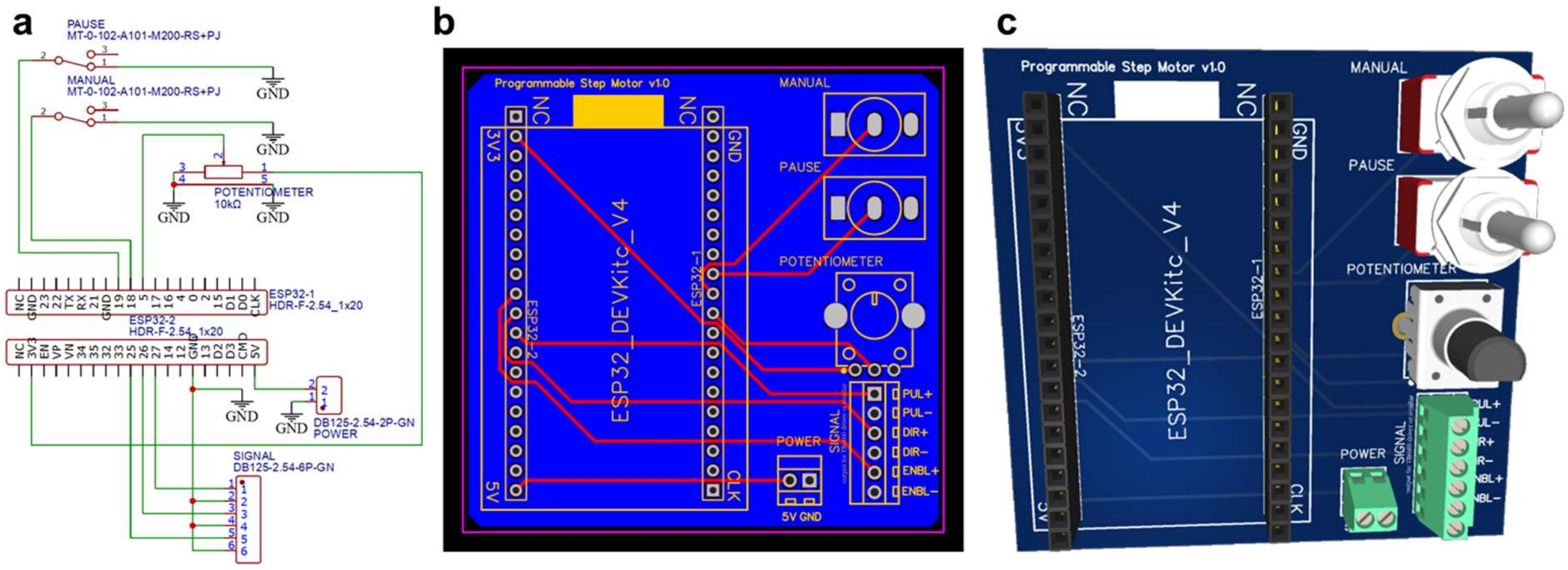

**Figure S1-1. PCB design and assembly for the rotation control system. a-b**, Circuit layout and **c**, schematic representation of the board.

The green connectors on the bottom right side (PUL, DIR and ENBL) should be wired to the TB6600 driver as follows: PUL+ from the PCB goes to PUL+ on the TB6600, while PUL- remains floating and is not connected in the PCB; DIR+ connects to DIR+, and DIR- to DIR-; both ENBL+ and ENBL- remain unconnected. Additionally, on the TB6600 side, PUL- and DIR− must be linked together with a short cable, leaving ENBL+ and ENBL- unconnected. See Figure S1-2 a for a quick view of the required connections.

The PCB is powered through a 5 V, 0.15 A power supply (e.g., a standard USB charger) via the green connector labelled “power” (see Figure S1-1c), while the TB6600 should be powered through a high-current switching power supply, such as 24V, 4.5A. The polarity of the power connections must be ensured, by connecting the negative or common wire of the power supply to GND and the other (usually labelled as VCC), to the pin with 5V and VCC in the PCB and TB6600, respectively.

The remaining connections of the TB6600 (A+, A-. B+ and B-) should be connected to the NEMO 23 200 stepper motor as follows: A+ to the red cable (STEP B in the manual), A- to the blue (STEP D), B+ to the black (STEP A), and B- to the green (STEP C), as shown in Figure S1-2 a. The SW1–SW6 switches on the TB6600 driver are used to set the motor current and the microstepping resolution. In our system, the drivers are configured to provide a sufficiently high microstep resolution, ensuring the required rotational precision (for further information, refer to the TB6600 manual). Specifically, the switch settings are: SW1 OFF, SW2 ON, SW3 OFF, SW4 ON, SW5 OFF, and SW6 ON. Figure S1-2b illustrates the placement of the power supply and the TB6600 driver inside a custom-built enclosure (a simple box with drilled holes to fix the elements).

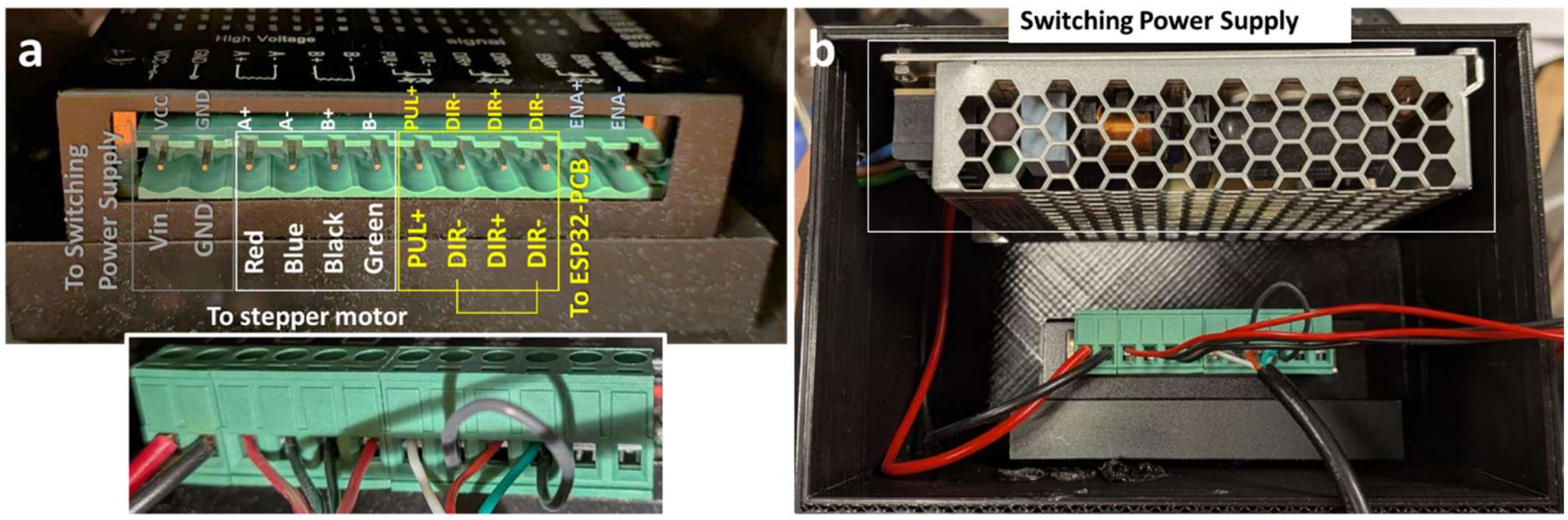

**Figure S1-2. TBA6600 connections to the ESP32 and stepper motor. a,** Shows the wiring layout, indicating the connections from the TB6600 driver to the power supply (grey), the stepper motor (white), and the ESP32-PCB (yellow). For clarity, the labels on the TB6600 have been written on top of the squares. **b,** Photograph illustrating the placement of the power supply and the TB6600 driver inside a custom-built enclosure.

2. **Mechanical connection from the motor to the transfer bar.** To efficiently transmit rotation, the stepper motor must be positioned close to the transfer bar. The required length of the synchronous belt depends on the distance between the motor and the transfer bar. In our configuration, the motor is mounted vertically above the bar, at approximately 27 cm distance, resulting in a belt circumference of 69 cm. The motor is mounted on top of an aluminum profile fixed to the vacuum frame structure, ensuring stability during operation. The motor is secured to this profile using a custom-made mounting bracket (Figure S1-3a, bottom), which is provided as an STL file in Supplementary Files S1 and can be fabricated using an FDM 3D printer (printed in our case with PLA). The motor is fastened to the bracket with four M4 screws and nuts (see Figure S1-3 a).

A GT2 timing wheel is coupled to the motor shaft to transmit rotation via a rubber synchronous belt. The custom bracket allows fine adjustment of the motor position, ensuring proper belt alignment and enabling tensioning by applying a small lever. After adjustment, the bracket is locked in place using the profile screw, resulting in a rigid assembly with the belt properly tensioned, as shown in Figures S1-3 b-d. It is important to note that the transfer bar must remain firmly fixed to prevent bending or misalignment after the belt coupling; this is achieved by means of threaded rods and small aluminum angle brackets, which maintain a fixed motor-to-bar distance while simultaneously suppressing bar deflection, as indicated in Figure S1-3 d.

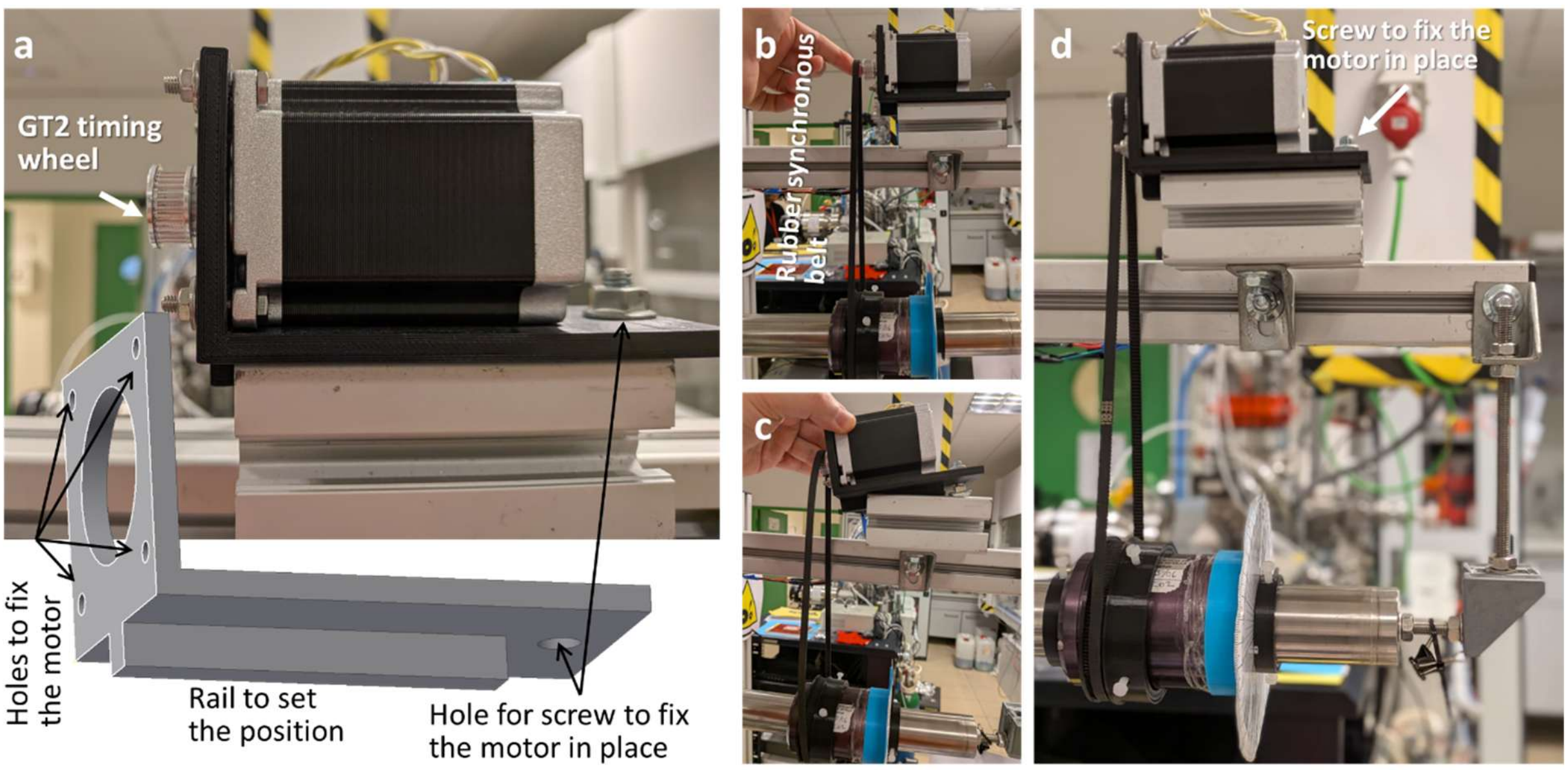

**Figure S1-3. Mechanical connection between the stepper motor and the transfer bar. a,** Detail of the motor mounted on a custom-made FDM printed bracket, showing the GT2 timing wheel and the bracket design with holes for motor fixation and a rail for positional adjustment. **b-c,** Installation of the rubber synchronous belt connecting the motor to the transfer bar. **d,** Final assembly with the motor secured in place and the belt properly tensioned. The bracket allows fine adjustment of motor position and tilt to ensure correct belt alignment.

Several custom-made components, fabricated using an FDM 3D printer, are mounted on the transfer bar to enable rotation and angle visualization. The first component, shown in Figure S1-4 a, incorporates GT2 gears for coupling the rubber synchronous belt. This part features internal hexagonal holes designed to hold M4 nuts, while M4 plastic screws (in our case, made of Teflon) are inserted from the opposite side. A total of four screws secures the component firmly to the transfer bar. The piece consists of two interlocking parts, as indicated in Figure S1-4 a, with the second part acting as a guide to prevent the belt from slipping. The fully assembled system is shown in Figure S1-4 b. The diameter ratio between the motor's timing wheel and the CAD-designed gear piece is 5:1 (16 mm vs. 80 mm). Since the stepper motor has a resolution of 1.8° per step, the resulting angular resolution on the larger wheel is 0.36° per step. If higher precision is required, this can be achieved by increasing the diameter of the printed gear piece and/or by selecting a stepper motor with a finer step resolution.
Additionally, a flat disk is attached to the rear side of the transfer bar (positioned and glued), together with a retaining ring with angle markers. The retaining ring uses the same fixation mechanism as the gears for the transfer bar, incorporating hexagonal holes for M4 metallic nuts and plastic screws (see Figure S1-4 c-d). This arrangement enables angle visualization when a simple goniometer (printed on paper and laminated) is placed on the flat disk, as shown in Figure S1-4 d. Figure S1-4c displays all custom components, including a second retaining ring mounted at the front to lock the bar in position and prevent any movement during rotation.

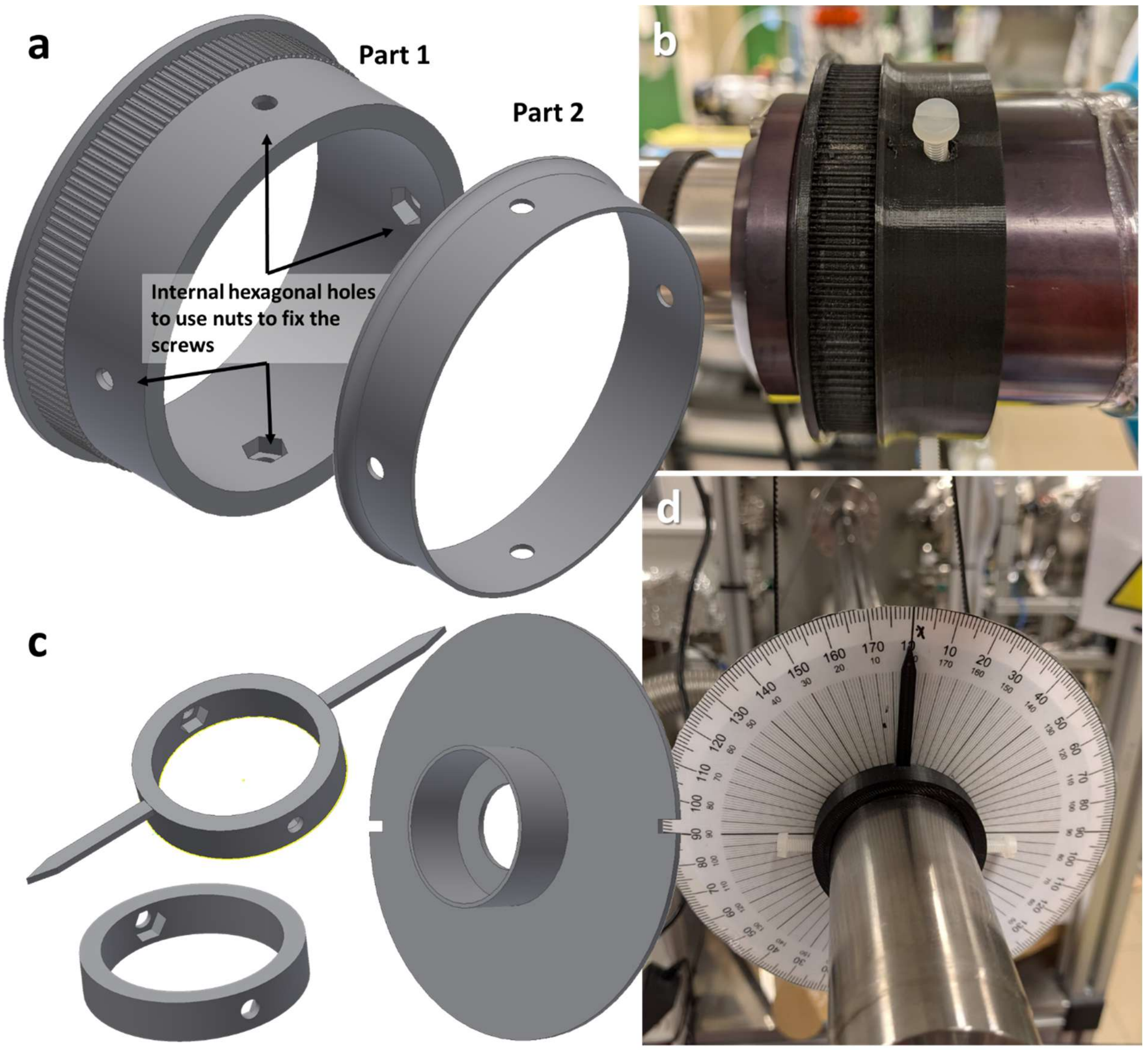

**Figure S1-4. Custom components attached to the transfer bar. a,** CAD design of the two-part component that holds the GT2 timing gears for coupling the rubber synchronous belt. The inner part includes hexagonal holes for M4 nuts, enabling secure fixation with plastic screws. **b,** Assembled component mounted on the transfer bar. **c,** Additional custom pieces, including a flat rear disk and retaining rings designed to lock the bar and prevent axial movement. **d,** Final setup showing the laminated paper goniometer positioned on the rear plate for angle visualization during rotation.

3. **Software interface and instructions to use the PCB and software.** To enable communication between the PCB and the stepper motor, the control program must be uploaded to the ESP32 via a PC. The source code is provided in Supplementary Files S1 within the .zip file.

Start by installing the Arduino IDE from https://www.arduino.cc/en/software. Once installed, open Preferences and add the following URL in the Additional Board Manager URLs field: https://dl.espressif.com/dl/package_esp32_index.json

Then go to Tools > Board > Board Manager, search for ESP32, and install the package from Espressif Systems. After installation, select ESP32 Dev Module (or the specific variant you are using) under Tools > Board, and choose the correct USB port under Tools > Port.

Connect the ESP32 to your computer via USB. Extract the .zip file from Supplementary Files S1 and open the .ino file in Arduino IDE. If the code requires additional libraries (e.g., for Bluetooth or Wi-Fi), install them via Sketch > Include Library > Manage Libraries. Click Verify to compile the code and then Upload to flash the program to the ESP32. After flashing the code, the program is in the ESP32 memory, and no further connection to the PC is required.

Once the ESP32 is powered, it can be accessed remotely via Bluetooth using any serial communication app available in mobile app stores (e.g., Serial Bluetooth Terminal for Android or similar apps for iOS). To connect, open the app and search for the ESP32 device in the list of available Bluetooth terminals. The device name usually appears as *StepMotor_Number* (as defined in the provided code). Select the device and establish the connection. Keep the Bluetooth terminal app configured to send a newline on Enter. If your app offers options like CR, LF, or CRLF, choose one that matches Arduino Serial readString behavior; LF usually works.

The following shows a typical workflow to define the rotation of the bar:

a) **Communicate with the PCB/ESP32.** First, power and wire the PCB and the TB6600 as described earlier. Open the serial Bluetooth app, search for the device name *StepMotor_Number*, and press "connect". After pairing, you can send single-letter commands and parameters to the ESP32 through the app's terminal interface. To display the list of available commands, simply type "h <enter>". The ESP32 will return a help menu showing all supported commands for rotation control and calibration.

b) **Calibrate steps per revolution**. We recommend calibrating every time the motor and belt have been connected or disconnected, as this helps detect mechanical issues. The PCB uses a potentiometer to define the rotation speed during calibration. First, switch to manual mode (upper switch in Figure S1-1 b–c) and adjust the potentiometer so the rotation is very slow or stopped (if possible). Then type "C <enter>", to enter Calibration Mode and follow the on-screen instructions.

The system requires a full turn of the transfer bar (which can be checked using the goniometer described in Section 2) for calibration. Increase speed gradually by adjusting the potentiometer (see Figure S1-1 b-c), keeping the same rotation direction (clockwise or anticlockwise) throughout calibration. This direction determines the rotation sign for future operations, so always perform calibration using the same rotation direction by adjusting the potentiometer consistently clockwise or anticlockwise. As approaching one full turn, reduce speed with the potentiometer for precision. Switch back to automatic mode (using the PCB switch) at the exact moment when the bar completes one full rotation. The system will automatically finalize calibration and store the number of steps for one full turn (the system sums all the steps accumulated). Record this number (which is displayed in the serial connection App) for future reference; it can help detect problems. Based on our experience, calibration error from calibrations on different days is typically below 0.2%.

If calibration is not required, you can verify the stored value by typing “S, <enter>”.

c) **Enter Programming Mode**. After calibration, you can enter rotation programming mode by typing “P, <enter>”. The program will ask for pairs of values: angle in degrees and duration in minutes. Example sequence:

“90, <enter> , 2, <enter>, -45, <enter>, 1, <enter>, <enter>”

This will rotate a total of +90° at constant speed for 2 minutes (angular velocity = +45°/min), then -45° (-45º/min) for 1 minute, and repeat the sequence. Finishing with an empty angle input exits Programming Mode. The device will compute the number of steps per segment and repeat the sequence.

The system also allows discrete rotations. For example, if you need a +90° discrete rotation, hold for 1 minute, then perform another discrete +90° rotation, type:

“90, <enter> , 0.1, <enter>, 0, <enter>, 1, <enter>, <enter> “

This rotates +90° very quickly (0.1 min=6 s) and then holds for 1 minute. Note that the maximum speed, and thus the minimum time for a full turn, depends on the TB6600 switch settings (SW1-6 in Section 1). In our configuration, the minimum is about 0.4 min (24 s). If very fast rotations are required, test the minimum time first; otherwise, incorrect rotations may occur.

You can also check the current steps per revolution stored in memory with the command “S, <enter>”. At any time, you can pause or resume using the PCB switch (middle switch labelled “pause” in Figure S1-1 b–c).

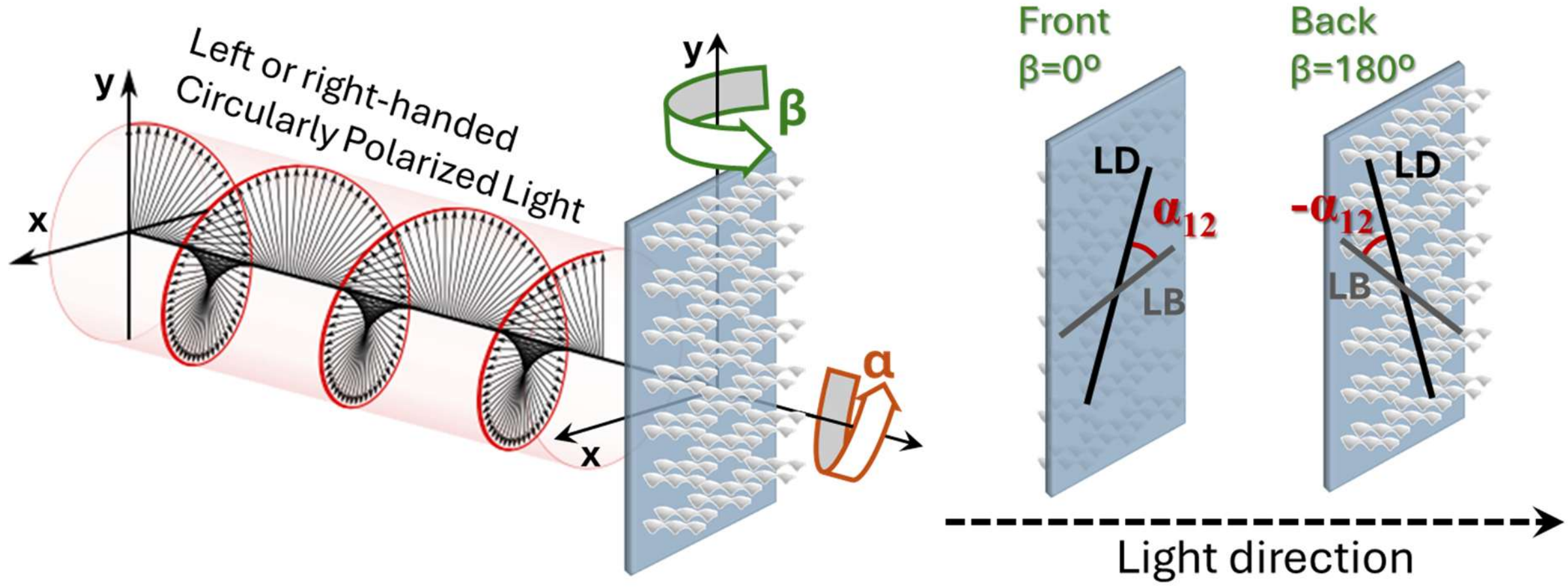

**Supplementary Figure S2. Schematic of the front-back averaging procedure.** The geometry of the measurement is illustrated through the rotation angles α and β. In linearly anisotropic samples, the measured CD includes a mixed linear term proportional to (LD'·LB - LD·LB'). Since this term reverses sign when the sample is flipped by β=180° relative to the light propagation direction, the $\theta_{chirop}$ component can be isolated by calculating the average of the spectra recorded from opposite orientations (front and back, β=0 and 180º).

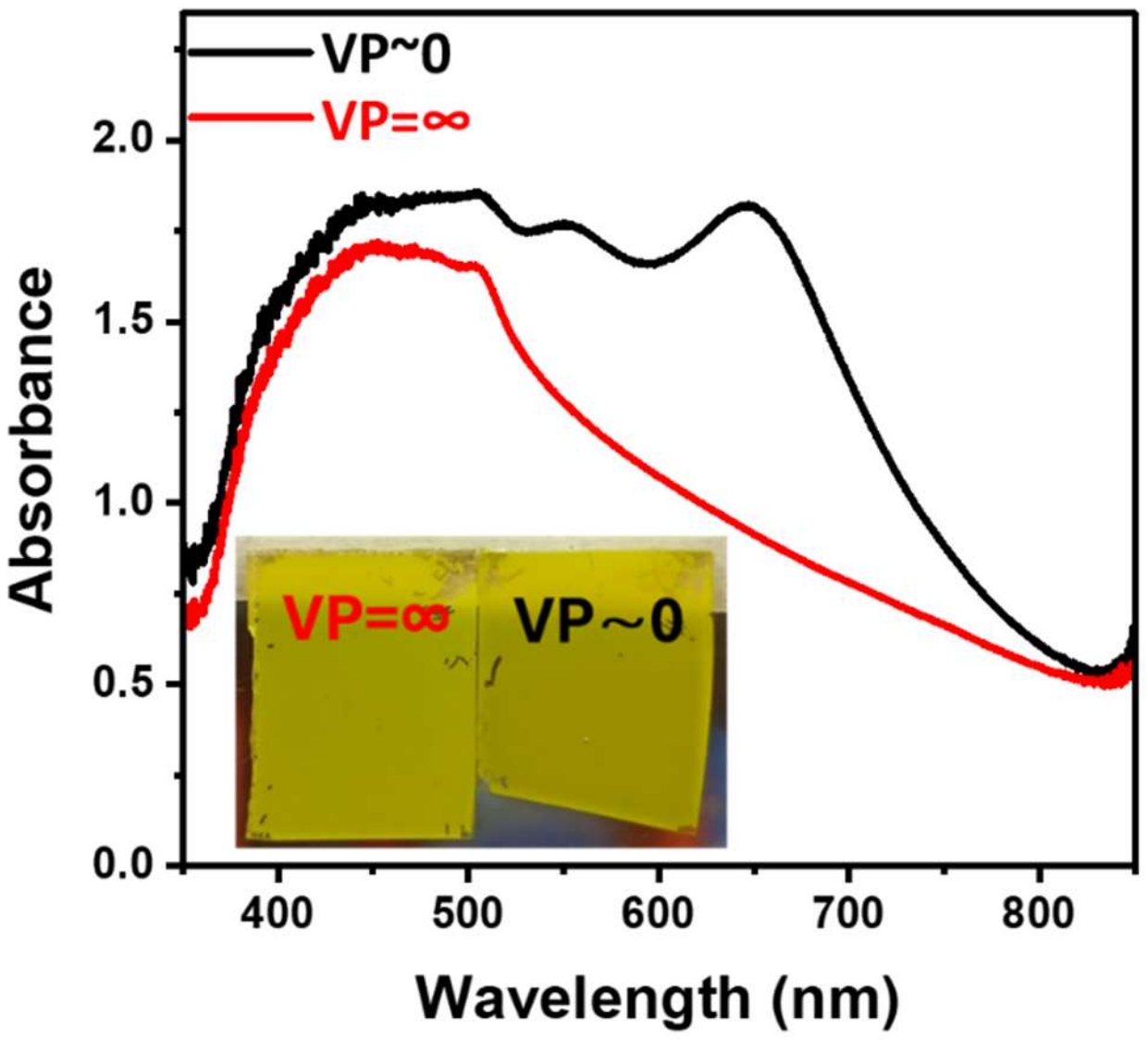

**Supplementary Figure S3. Absorbance spectra for the reference extreme cases, VP~0 and VP=∞.** The VP~0 sample shows a broader spectrum with additional features around 550 nm and 650 nm. Although the origin of this broadening is not clear, we hypothesize that it is related to the crystalline torsion induced during growth (with an approximate VP of ~8 nm), which presents measurable circular dichroism, albeit with amplitudes much lower than those observed for larger pitches. The inset shows a photograph of the samples.

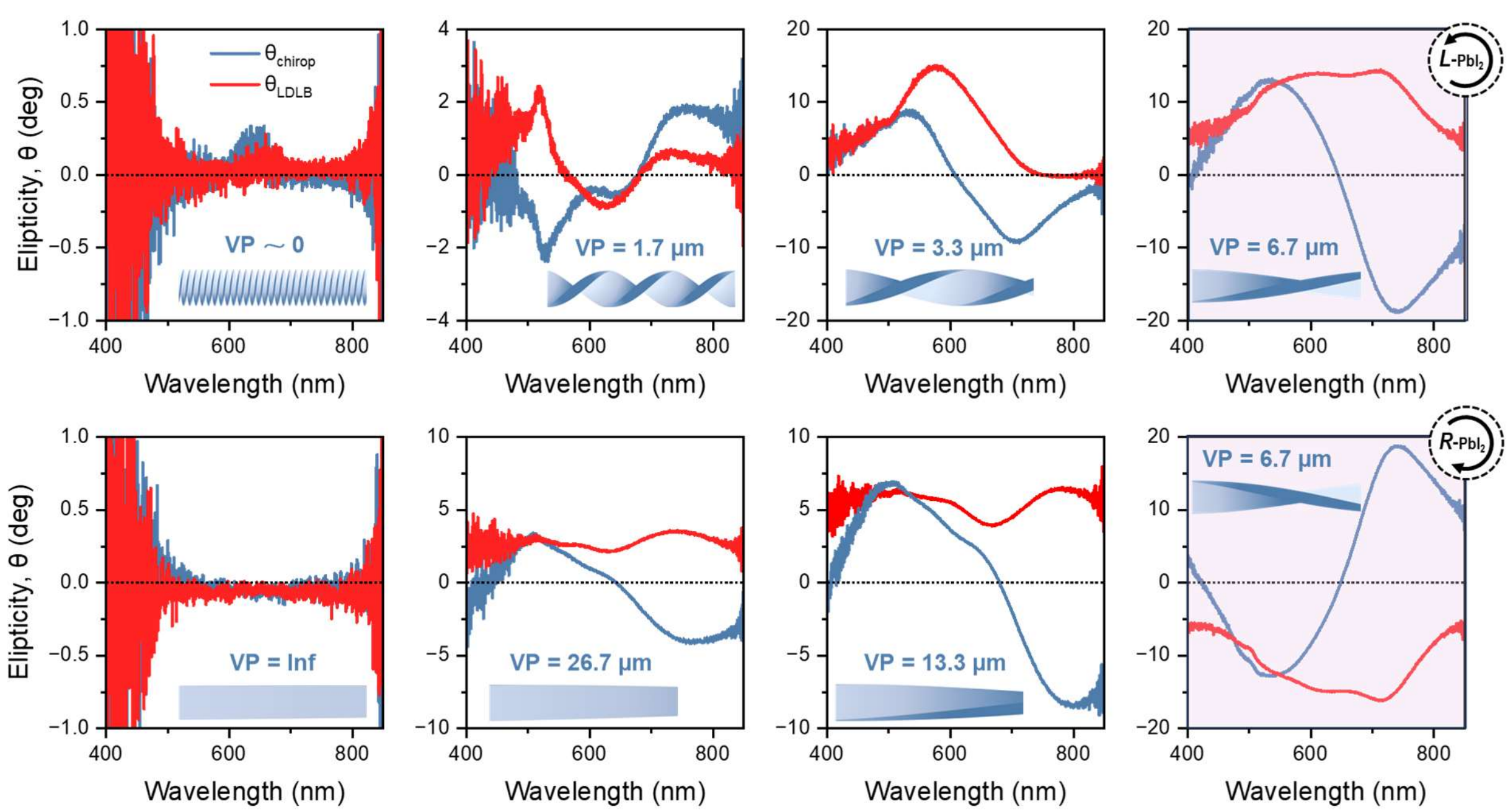

**Supplementary Figure S4. Comparison between Chiroptic Ellipticity ($\theta_{chirop}$) and LDLB ($\theta_{LDLB}$) terms**. The $\theta_{chirop}$ and $\theta_{LDLB}$ spectra exhibit maximum values for VP = 3.3-13.3 μm and negligible signal for VP=∞. All samples shown correspond to counterclockwise substrate rotation (*L*-$PbI_2$). For comparison, the spectra of the VP = 6.7 μm sample prepared using clockwise rotation (*R*-$PbI_2$) are also included in the bottom right panel, highlighting the inversion of the LDLB response upon reversal of the growth handedness.

**Supplementary Information S5. Circular Bragg Phenomenon**

In conventional GLAD spirals, the helical morphology of the nanostructure gives rise to the circular Bragg phenomenon, resulting in selective reflection and transmission of circularly polarized light within the Bragg regime.[1] The central resonance wavelength $\lambda_{Bragg}$ at which the spirals produce the selective reflection or transmission satisfies $\lambda_{Bragg} \approx n_{eff} \cdot VP$ , where $n_{eff}$ is the effective refractive index of the layer. This photonic bandgap phenomenon typically requires multiple full turns and sufficient thickness to act as a circular polarization filter. As Lakhtakia *et al.* emphasize, strong circular dichroism and high reflectance in the Bragg regime only emerge when the accumulated rotation spans many full turns (typically tens of periods) along the thickness direction.[2]

In contrast, our GLAD $PbI_2$ nanostructured layer exhibits giant CD values in the visible range (500–800 nm) despite geometries that are fundamentally incompatible with Bragg reflection. For instance, the configuration delivering the strongest CD corresponds to a pitch of VP=6.7 µm, yet the total film thickness of only 2.5 µm results in an accumulated twist of merely ~135°, far from even a single full rotation. Considering the high porosity of the nanostructures, the effective refractive index is expected to be substantially lower than that of bulk $PbI_2$ ($n > 3$ in the visible range). Even assuming a conservative effective refractive index of $n_{eff} \approx$ 1.5-2, the Bragg wavelength associated with a pitch of VP = 6.7 µm would still be located far beyond the visible region, at wavelengths exceeding 10-13 µm. Moreover, a circular Bragg mechanism would predict an approximately linear scaling of the spectral response with pitch. Consequently, increasing VP from 6.7 to 13.3 µm should shift the optical feature by roughly a factor of two. Experimentally, however, the chiroptical maximum evolves only from ~740 nm to ~800 nm, demonstrating that the observed response does not follow the behaviour expected for a Bragg reflection process.

Literature further supports that the circular Bragg phenomenon critically depends on a continuous helicoidal axis and structural periodicity. In GLAD, azimuthal rotation sculpts oblique columns into helices, and the degree of helicoidal twist strongly depends on the initial column inclination: highly tilted columns yield tight spirals, whereas nearly vertical columns produce very loose helices or remain essentially straight, eliminating the structural periodicity required for the Bragg regime.[3,4] When this helicoidal axis is absent or poorly defined, as occurs when columns are quasi-vertical, the Bragg regime collapses, and the material no longer exhibits circular-polarization selectivity.[2,5] Since our nanostructures grow quasi-vertically (~10º from the substrate normal, see Figure 1b in the main manuscript) and accumulate only ~135° rotation over 2.5 µm thickness for VP=6.7 µm, they lack the continuous helicoidal axis required for Bragg reflection, reinforcing that the observed giant CD arises from a different mechanism.

Therefore, the intense chiroptical response observed cannot originate from the circular Bragg effect but must arise from intrinsic material chirality and excitonic interactions. Unlike conventional helicoidal GLAD structures, where circular polarization selectivity emerges from periodic morphological spirals, the optical activity reported here originates from growth-controlled crystallographic torsion.

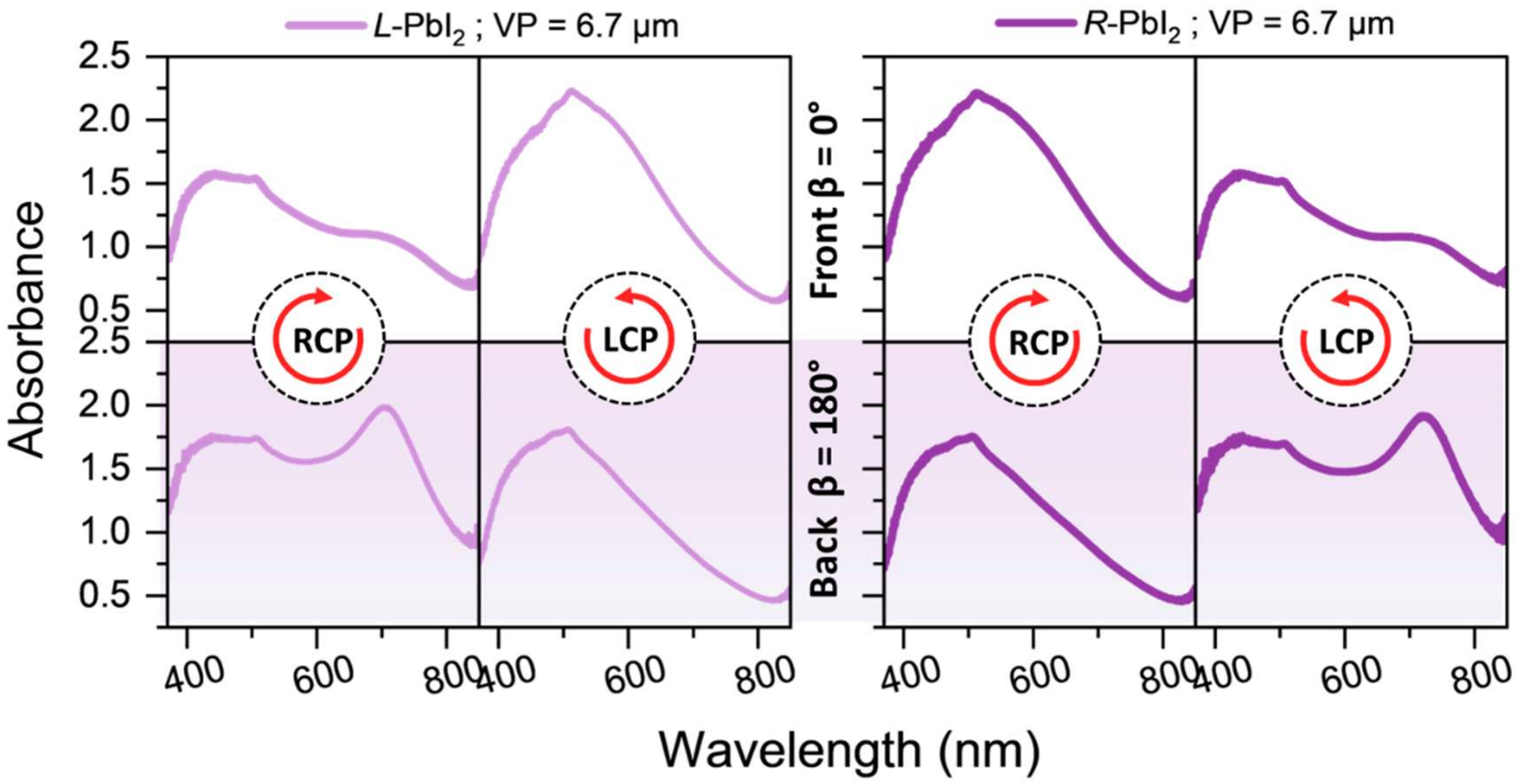

**Supplementary Figure S6. Front- and back-side chiroptical response of *L*- and *R*-$PbI_2$ nanostructures.** Absorbance spectra of $PbI_2$ with VP=6.7 μm for RCP and LCP light under front (β=0º, top) and back (β=180º, bottom) configurations for both "enantiomeric" samples, *L*-$PbI_2$ (left panels, substrates rotated counterclockwise), and *R*-$PbI_2$ (right panels, substrates rotated clockwise).

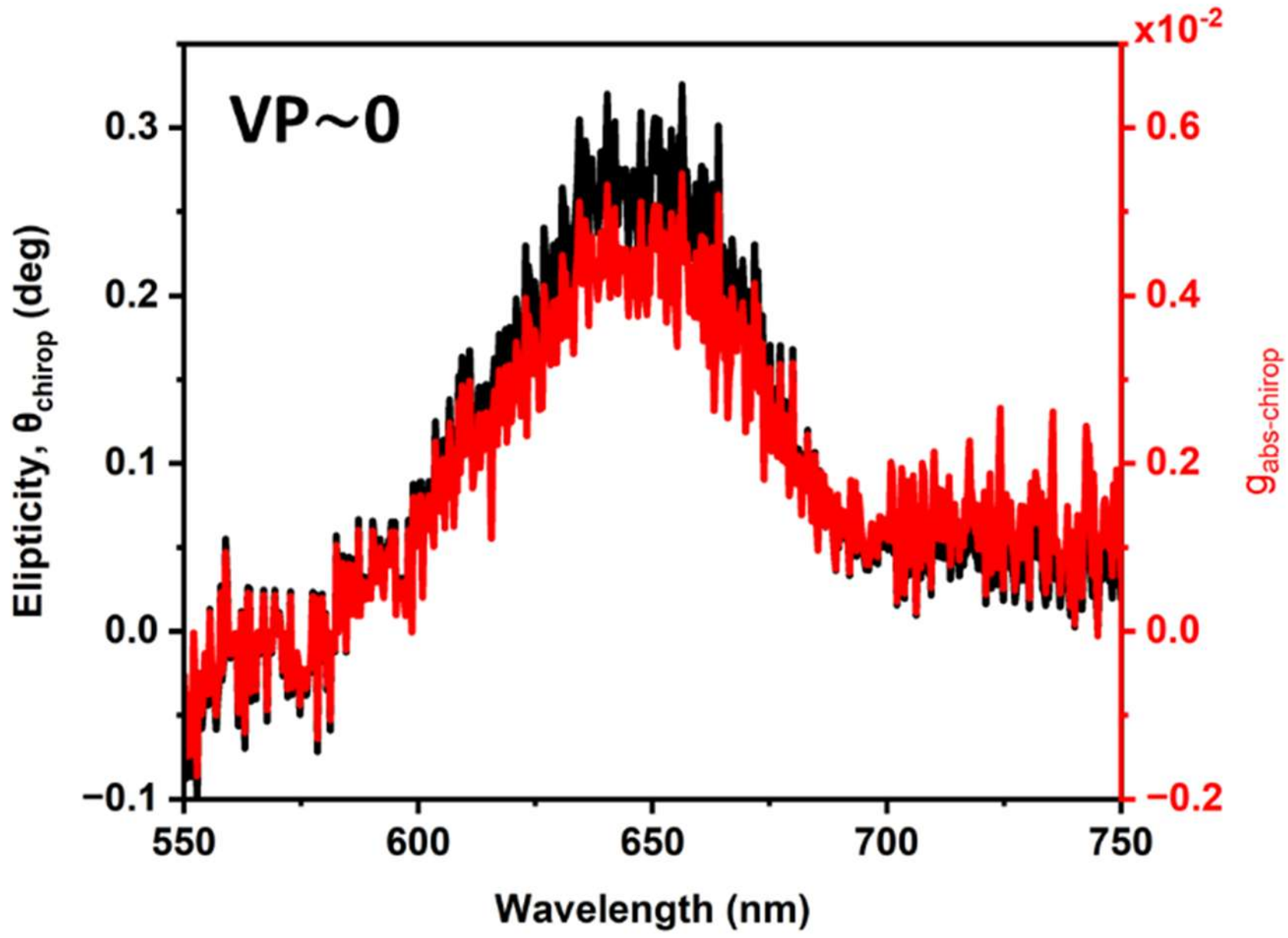

**Supplementary Figure S7. Chiroptical response in the rapid-rotation limit.** Ellipticity ($\theta_{chirop}$) and Disymmetry factor ($g_{abs\text{-}chirop}$ ) for VP~0 (VP=8 nm).

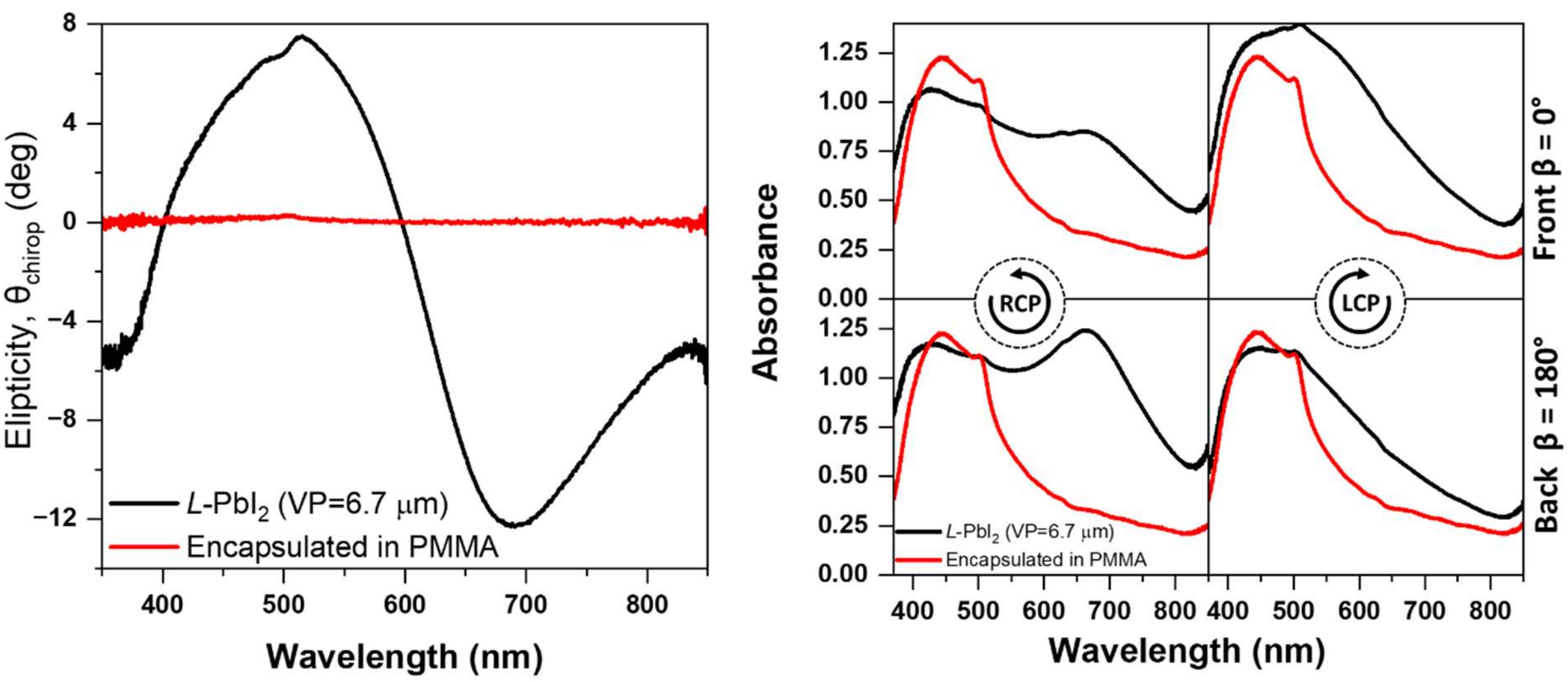

**Supplementary Figure S8. Collapse of the chiroptical response after PMMA encapsulation.** Standard PMMA spin-coating was tested as a protective layer, but the wet process proved unsuitable: exposure to liquid drastically reduced the chiroptical response. We attribute this loss to the surface-tension forces arising during wet processing, which induce capillary-driven collapse of the nanowalls. As the twisted nanoribbons lose their nearly vertical orientation, the chiral axis becomes increasingly misaligned with respect to the film normal, strongly reducing the observed optical activity. Although a residual chiroptical signal is still detectable, the reduction in intensity is so severe that any liquid-based approach to encapsulation or conversion becomes impractical.

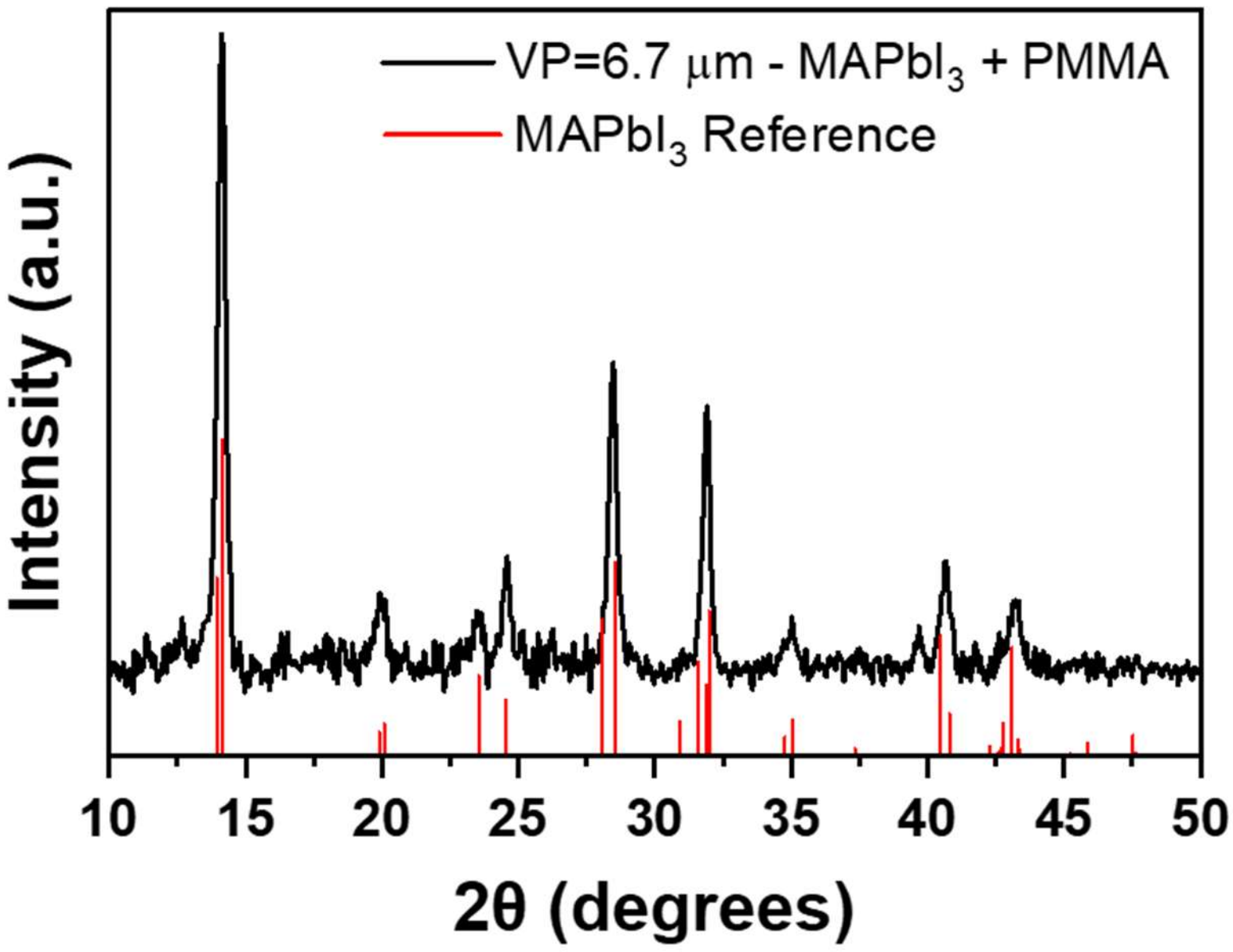

**Supplementary Figure S9. XRD confirmation of complete $PbI_2$-to-$MAPbI_3$ conversion.** XRD pattern of the VP = 6.7 μm *L*-$MAPbI_3$ sample after transformation by exposure to MAI vapours in a high-vacuum chamber. The diffraction pattern shows no trace of the $PbI_2$ (001) peak, typically observed at 12.7°, confirming complete conversion of $PbI_2$ into $MAPbI_3$. The sample was encapsulated with PMMA inside the glovebox prior to XRD measurements to maintain stability during exposure to ambient conditions.

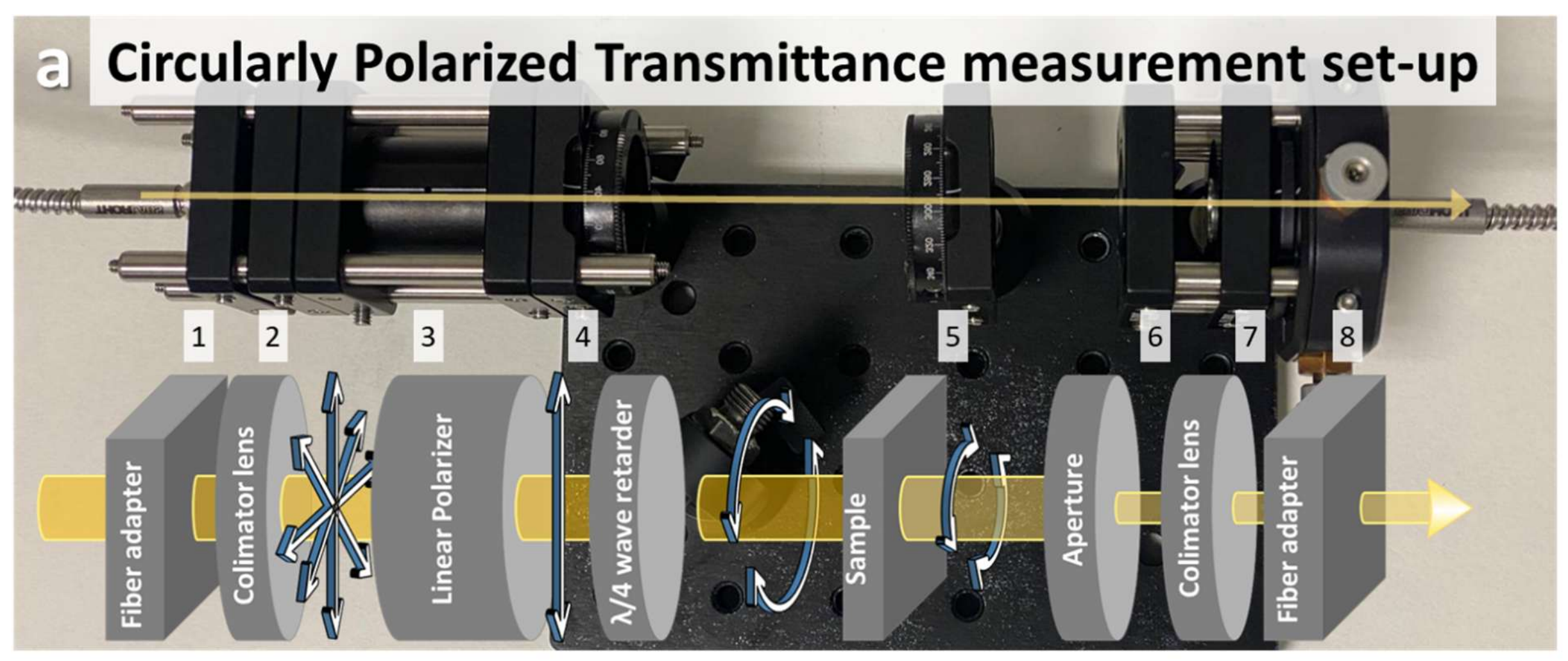

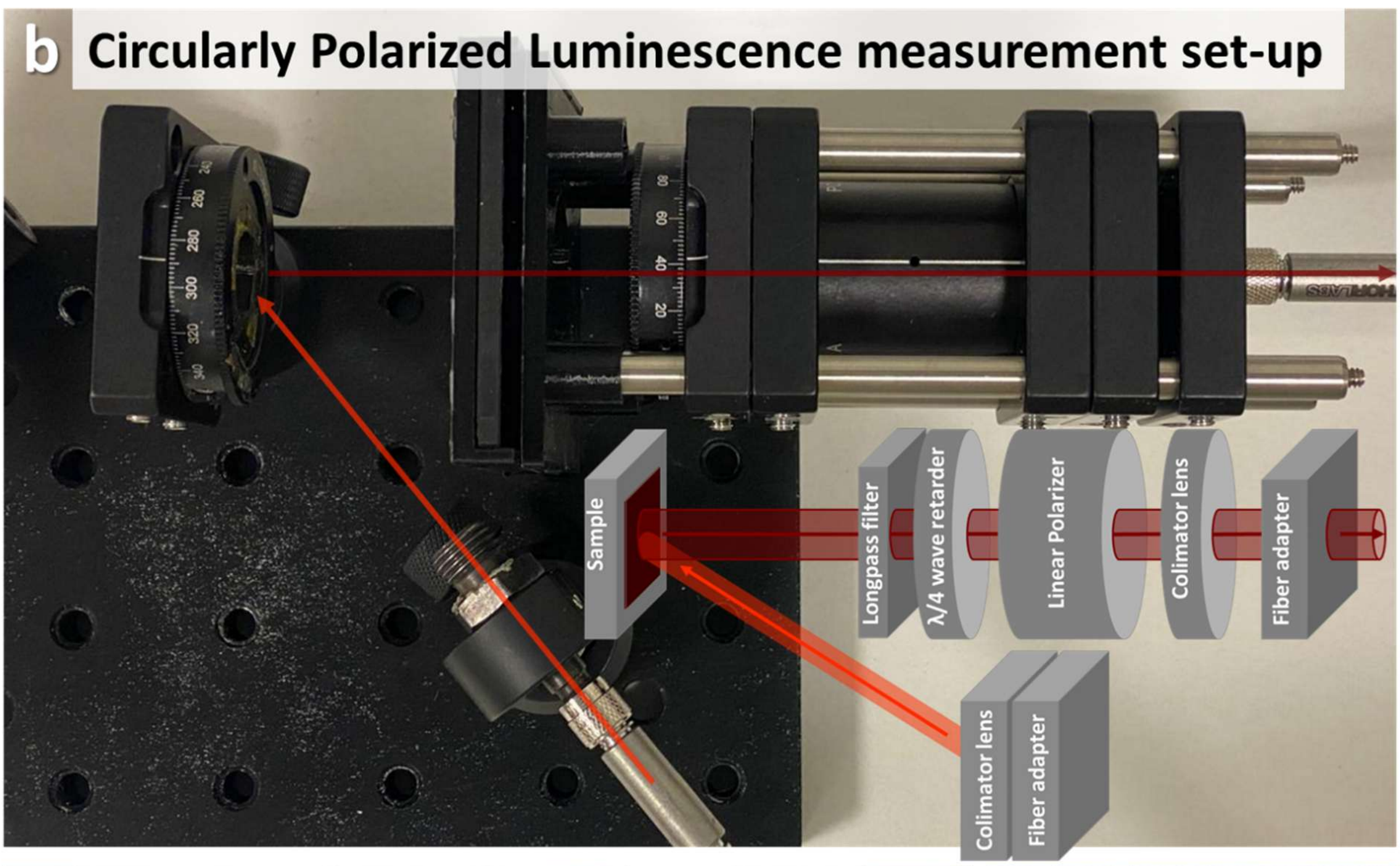

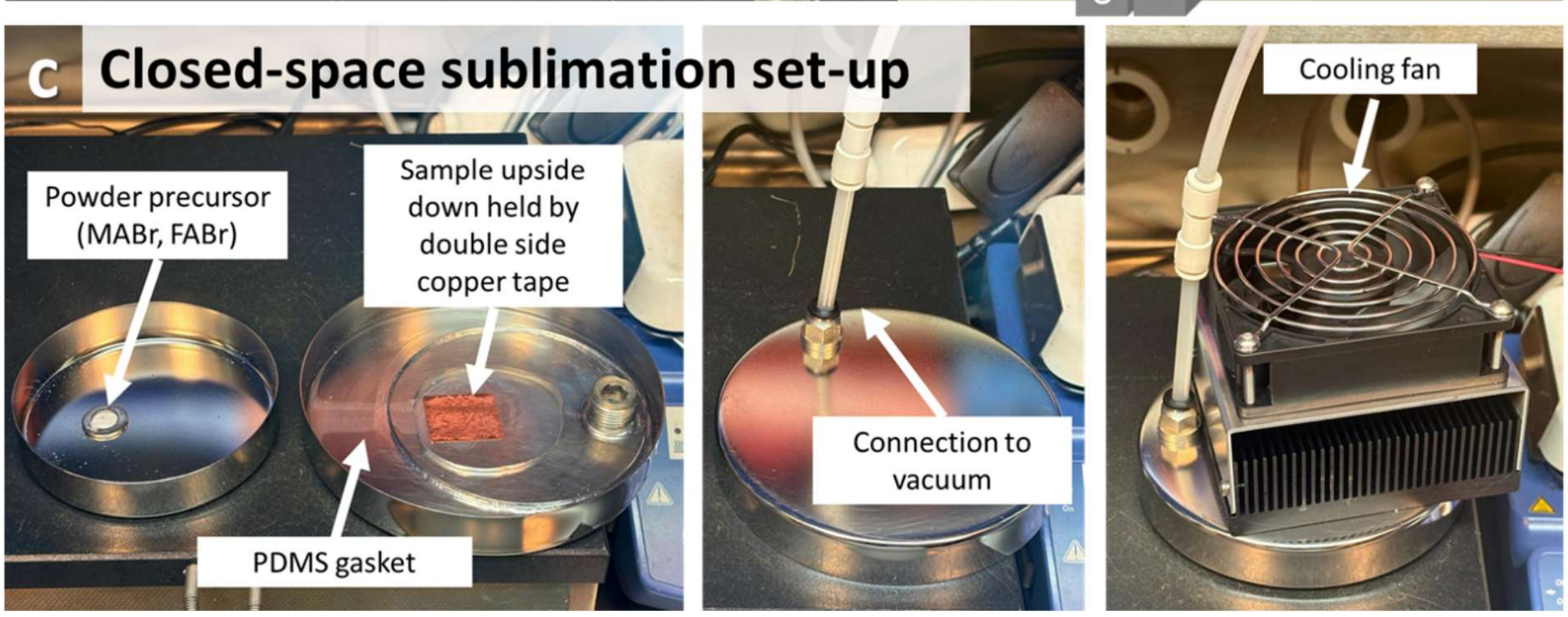

**Supplementary Figure S10. Experimental platforms for chiroptical measurements and closed-space sublimation. a-b**, Setups for circularly polarized light transmittance (a) and circularly polarized photoluminescence (b) measurements, both located inside the glovebox and coupled through optical fibres. **c**, Photographs of the closed-space sublimation system.

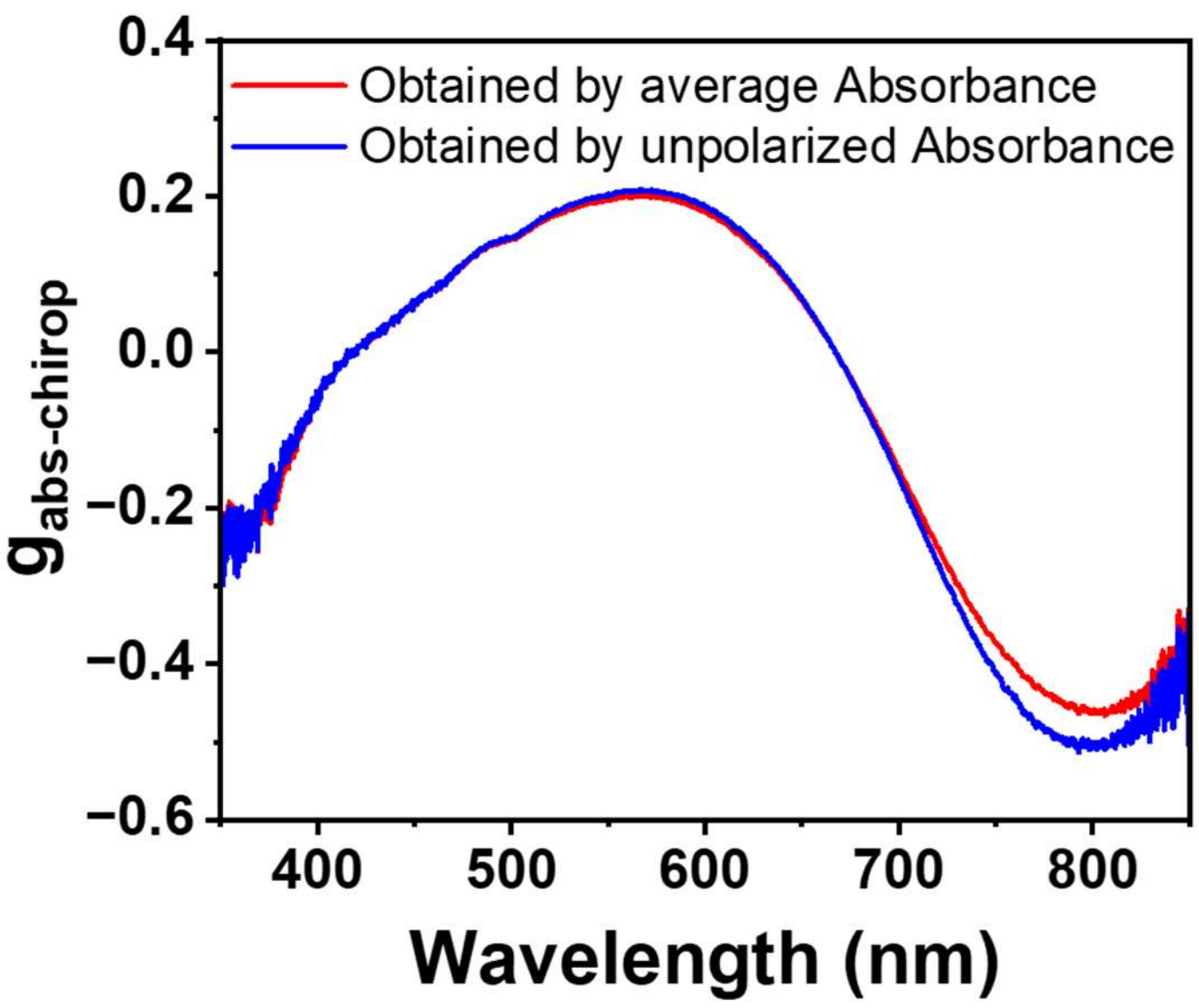

**Supplementary Figure S11. Validation of the conservative approach used for $g_{abs}$ estimation.** Comparison of $g_{abs\text{-}chirop}$ values obtained using absorbance estimated from the average of the front-side, back-side, and azimuthal measurements employed for chiroptical analysis, and using directly measured unpolarized absorbance. The comparison reveals deviations at longer wavelengths. Using the unpolarized absorbance yields systematically larger $g_{abs\text{-}chirop}$ values, confirming that the averaging procedure provides conservative estimates of the dissymmetry factor.